\documentclass[twoside]{article}
\usepackage{fleqn,espcrc2}
\usepackage{epsfig}

\usepackage{graphicx}

\newcommand{\AmS}{{\protect\the\textfont2
  A\kern-.1667em\lower.5ex\hbox{M}\kern-.125emS}}

\def\eg{{\it e.g.}}
\def\ie{{\it i.e.}}
\def\etal{{\it et al.}}
\def\etc{{\it etc.}}

\def\bi{\begin{itemize}}
\def\ei{\end{itemize}}
\def\ib{\item}
\def\BR{{\cal B}}

\def\epsbar{\bar{\epsilon}}

\def\nutau{\nu_\tau}
\def\DD{\displaystyle}
\def\lapprox{\sim\kern-1em\raise 0.6ex\hbox{$<$}}
\def\rapprox{\sim\kern-1em\raise 0.6ex\hbox{$>$}}
\def\ssection#1{\subsection{#1}}
\def\ben{\begin{equation}}
\def\een{\end{equation}}

\def\bmath{} 

\title{Semi-hadronic Tau Decays at CLEO}

\author{Alan. J.~Weinstein
        \address[MCSD]{California Institute of Technology, \\
        Pasadena, CA 91125, USA \\
        Representing the CLEO Collaboration}
        \thanks{Work supported by the US Department of Energy 
                and National Science Foundation.}
       }

\begin{document}

\begin{abstract}
We describe recent results from the CLEO experiment
on semi-hadronic decays of the tau lepton.
We discuss the analysis of sub-structure in the decays
$\tau^- \to \pi^-\pi^0 \nu_\tau$,
$(3\pi)^-   \nu_\tau$,
$(4\pi)^-   \nu_\tau$,
$(6\pi)^-   \nu_\tau$,
and $\eta X^-   \nu_\tau$.
Various applications of these results are also discussed.
\vspace{1pc}
\end{abstract}

\maketitle

This contribution reviews several new or updated results on
substructure in semi-hadronic tau decays, recently published by the
CLEO Collaboration.  These results are all based on data collected
with the CLEO detector at Cornell's CESR collider, in the reaction
$e^+e^-\to \tau^+\tau^-$ at $\sqrt{s} \simeq 10.6$ GeV.  Most of these
results are based on $\approx 4.3\times 10^6$ $\tau^+\tau^-$ pairs
collected with the CLEO II detector between 1990 --- 1995. Some (where
noted) also use $\approx 8.0\times 10^6$ $\tau^+\tau^-$ pairs
collected with the CLEO II.V detector (which includes a silicon vertex
detector and better drift chamber tracking) between 1995 --- 1999.


\section{Hadronic substructure of tau decays}

All the tau decay branching fractions larger than 1\%\
have been measured reasonably well,
with errors that are dominated by systematic uncertainties.
The next step in exploiting tau decays to learn more about
the Standard Model is to explore the substructure of the
decays to three or more final state particles.

For the leptonic decays $\tau\to \ell\bar{\nu}_\ell\nu_\tau$,
the substructure is parameterized by the Michel parameters;
precision measurements of these serve to constrain 
the charged weak couplings of the tau, beyond the
well-understood  Standard Model $V-A$ couplings.

For the semi-hadronic decays $\tau \to X\nu_\tau$,
the study of hadronic substructure is
a clean probe of one of the least well understood
aspects of the Standard Model: low energy meson dynamics.

\begin{figure}[htb]
\begin{center}
\centerline{\psfig{figure=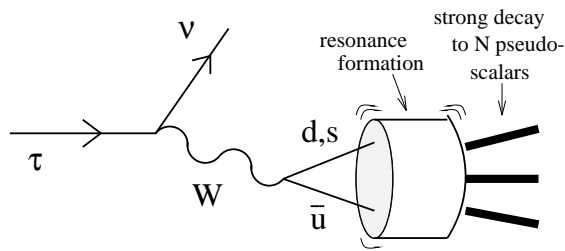,width=75mm}}
\caption[]{\label{fig:resonance}
Cartoon of semi-hadronic tau decay to mesonic final states
dominated by intermediate resonances.}
\end{center}
\end{figure}

In tau semi-hadronic decays, momentum transfers are small,
so final states are dominated by resonances (vector, axial-vector,
and to a lesser extent, scalar and tensor resonances);
see Fig.~\ref{fig:resonance}.
Lacking a fundamental theory of meson resonance dynamics,
these processes are described using models.

The weak decay $\tau^-\to W^{*-}\nu_\tau$ 
(and its charged conjugate, which is implicitly assumed
throughout this paper) is assumed to be well
described by the Standard Model $V-A$ current,
and the poorly-known hadronic physics in the transition
$W^{*-}\to X^-$, where $X$ is a system of hadrons,
is described by a spectral function $v(q^2)$ \cite{ref:tsai,ref:thacker},
where $q^2 \equiv M_{X}^2$ is the invariant mass squared of the 
hadronic system. The ``production'' and ``decay''
of the $X^-$ system separates cleanly;
for decay to non-strange final states, 
\begin{equation}
 { \frac{d\Gamma}{dq^2}}  = 
          \frac{G_F^2\, |V_{ud}|^2}{32\pi^2\, M_\tau^3}\, 
   { 
   (M_\tau^2-q^2)^2\,(M_\tau^2+2q^2)
   \, { v(q^2)}, }
\end{equation}
with an analogous expression for Cabibbo-suppressed decays.
The spectral function contains all the strong interaction dynamics.

CLEO can measure { $v(q^2)$} using exclusive final states
like $X = 2\pi$, $3\pi$, \etc; it is 
more problematic to do inclusive studies,
due to backgrounds and cross-feeds between different 
exclusive final states.

\ssection{Theory of tau semi-hadronic decays}

The low-energy dynamics of strongly-interacting mesons
is the poorest understood aspect of the Standard model.
The tools that we have to understand 
the structure of the spectral function 
{ $v(q^2)$} are:
Conservation laws (Lorentz invariance, isospin, $SU(3)_f$, $G$-parity, \etc);
resonance dominance models and the PDG catalog;
the Conserved Vector Current (CVC) constraints;
QCD sum rules (for inclusive studies);
Chiral perturbation theory of pseudoscalar mesons and 
higher mass resonances
(applicable only for momenta close to threshold);
QCD on the lattice;
and non-perturbative models inspired by S-matrix theory.

For { $\tau^- \to \nu_\tau \bar{u}d$ (\ie, with
strangeness $= 0$), the strong hadronization of the 
$\bar{u}d$ into observable mesons conserves parity, isospin, and G-parity,
so that $J^P$, $I^G$ are good quantum numbers of the
hadronic current from weak decays
\cite{ref:tsai,ref:thacker}.
The weak vector current produces systems of pions with $J^P = 0^+$
or $1^-$,
even G parity, and even numbers of pions.
The weak axial-vector current produces systems of pions with $J^P = 0^-$
or $1^+$,
odd  G parity, and odd  numbers of pions.

CVC relates the { vector} part of { $W^{*-}\to \bar{u}d$}
in, \eg, ${ \tau^-\to X_{had}^- \nu_\tau}$
to the { isovector $I=1$}
part of { $\gamma^*\to \bar{u}u, \bar{d}d$}
in \eg, { $e^+e^-\to X_{had}^0$}, with $s = q^2$:
\ben
{ \sigma^{(I=1)}_{e^+e^-\to X^0}(s)} = 
            \left(\frac{4\pi^2\alpha^2}{s}\right)\,
               { v_{X}(s)}.
\een
CVC also forbids { $J^P = 0^+$} final states from forming.
The axial-vector current is not so heavily constrained (PCAC, sum rules);
tau lepton decay is well suited for study of light axial vector mesons.

\section{$\tau^-\to \pi^-\pi^0\nu_\tau$}

We expect the $2\pi$ final state to be dominated by the vector resonances
$\rho(770)$, and its (broad and thus poorly understood)
radial excitations $\rho^{\prime}(1450)$, $\rho^{\prime\prime}(1700)$.
The masses, pole widths, and mass-dependent widths of these resonances
are of interest. It is also of interest to search for
unexpected (CVC-violating) scalar resonances, or non-resonant 
contributions with well-defined Lorentz structure.

The recently published results from CLEO \cite{ref:cleorho}
are based on the CLEO-II sample of 4.3$\times 10^6$ produced
tau pairs.
Approximately 87,000 events consistent with $\tau^\mp\to\pi^\mp\pi^0\nu_\tau$
were selected.
The $\pi\pi^0$ mass distribution is shown in Fig.~\ref{fig:cleorho}.

\begin{figure}[htb]
\centerline{
\hbox{\epsfxsize75mm\leavevmode\epsfbox[106 330 494 730]
     {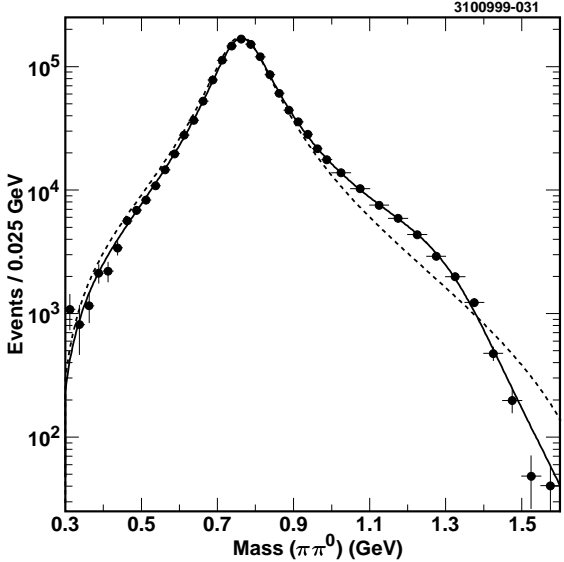}}
}
\caption[]{\label{fig:cleorho}
The $\pi\pi^0$ mass distribution, based on 87,000 events observed by CLEO.
The points are data, corrected for efficiency, background, and resolution
(via an unfold procedure).
The solid curve is a fit (described in the text) including 
contributions from the $\rho(770)$, $\rho(1450)$, $\rho(1700)$ resonances.
The dashed curve is a fit with the $\rho(770)$ lineshape only.
}
\end{figure}

This mass distribution (with $q^2=m^2(\pi\pi)$)
is modeled in terms of the spectral function:
\ben
   { v_{\pi\pi}(q^2) = \frac{1}{12\pi}\vert F_\pi(q^2)\vert^2
     \left(\frac{2p_\pi}{\sqrt{q^2}}\right)^3 } ,
\een
where $F_\pi(q^2)$ is the pion charged-current form-factor.
CVC predicts that this form factor should be the same
(up to isospin-violating effects) as the neutral current form factor
for $\gamma\to\pi^+\pi^-$,
whose value at $q^2=0$ is equal to 1.
We have allowed $F_\pi(0)$ to float in our fits,
although it has been argued \cite{ref:davier}
that a better approach is to fix it to its predicted value
and extrapolate to $q^2>4m_\pi^2$ in some smooth and well defined way.

\ssection{Model-dependent fits}

CLEO has used two phenomenological models of $F_\pi(q^2)$ to fit the
observed spectrum.
In the model of K\"uhn and Santamaria (K\&S, \cite{ref:KandS}),
$F_\pi(q^2)$ is a coherent sum of simple Breit-Wigner lineshapes:
\ben
        { F_\pi(q^2)} 
                \propto  BW_\rho + { \beta}\, BW_{\rho^\prime}
                        + { \gamma}\, BW_{\rho^{\prime\prime}} +\cdots,
\een
\ben
  BW_\rho = \frac{{ M_\rho}^2}
                 {({ M_\rho}^2 - q^2) 
                  - i \sqrt{q^2}\, { \Gamma_\rho(q^2)}}.
\een
The $q^2$-dependence of the width $\Gamma_\rho(q^2)$ 
is calculated assuming simple P-wave decay
into two pions, only. 

The model of Gounaris and Sakurai (G\&S, \cite{ref:GandS})
is somewhat more complicated, based on assumed effective range formula 
for the P-wave $\pi\pi$ scattering phase shift.

In both models,
the masses and pole widths of the resonances are free (fit) parameters.
Since there is negligible sensitivity to the pole mass and width
of the $\rho^{\prime\prime}$, they are fixed to be 1700 MeV and 235 MeV,
respectively.

The CLEO results favor the G\&S model over K\&S model.
The results of the fits are given in Table~\ref{tab:rhofit},
in the context of that model,
and are compared with analogous results from ALEPH (using $\tau\to\pi\pi\nu_\tau$)
\cite{ref:aleph2pi}
and a fit to $e^+e^-\to \pi^+\pi^-$ data in Ref.~\cite{ref:KandS}.

\begin{table}
\caption[]{\label{tab:rhofit}
Results of the fits to the CLEO $\tau\to \pi\pi\nu_\tau$ data
using the G\&S (ref.~\protect\cite{ref:GandS}) model.
The results are compared with analogous results from 
ALEPH (ref.~\protect\cite{ref:aleph2pi})
and from $e^+e^-$ data (ref.~\protect\cite{ref:KandS}).
}
{
\begin{center}
\begin{tabular}{c|ccc}
\hline\hline
                  & \multicolumn{3}{c}{G\&S Model} \\
                  & CLEO  & ALEPH & $e^+e^-$ \\
\hline
  $M_\rho$
                  & { $775.1\,(1.1)$} 
                  & {    $776.4\,(0.9)$} & 776 \\
  $\Gamma_\rho$
                  & { $150.4\,(1.4)$} 
                  & {    $150.5\,(1.6)$} & 151 \\
  $\beta$
                  & $-0.121\,(10)$ & $-0.077\,(8)$  & $-0.052$ \\
  $M_{\rho^\prime}$
                  & { $1406\,(15)$}   
                  & { $1400\,(16)$}   & 1330 \\
  $\Gamma_{\rho^\prime}$
                  & $455\,(41)$    & $\equiv 310$   & $270$ \\
  $\gamma$
                  & $0.032\,(9)$   & $ 0.001\,(9)$  & $-0.031$ \\
  $ | F_\pi(0) |^2$
                  & $1.03\,(2)$    & $\equiv 1$     & $\equiv 1$ \\
\hline
 $\chi^2$/dof 
                  & { 22.9/23}        
                  & {    54/65}          & 151/132 \\
\hline\hline
\end{tabular}
\end{center}
}
\end{table}

These fits yield rather precise values for the 
charged { $\rho(770)$} mass and width
(the $e^+e^-$ results are for the $\rho^0$).
The mass of the { $\rho(1450)$} is around
$\sim 1400$~MeV, but this is a very model-dependent result,
strongly influenced by the presence or absence of the $\rho(1700)$.

\ssection{Test of CVC with $|F_\pi(q^2)|^2$}

The results of the fit to the CLEO data can be compared directly
with the $e^+e^-$ data, as a test of CVC.
These comparisons are shown in Figs.~\ref{fig:fpi}
and \ref{fig:dfpi}.
In these figures, 
$|F_\pi(q^2)|^2$ is extracted directly from CLEO $\tau$ data.
The $I=0$ contribution from $\omega\to\pi^+\pi^-$,
including $\rho-\omega$ interference,
is removed from $e^+e^-$ data.

\begin{figure}[htb]
\centerline{\hbox{\epsfxsize75mm\leavevmode\epsfbox[106 336 494 730]
     {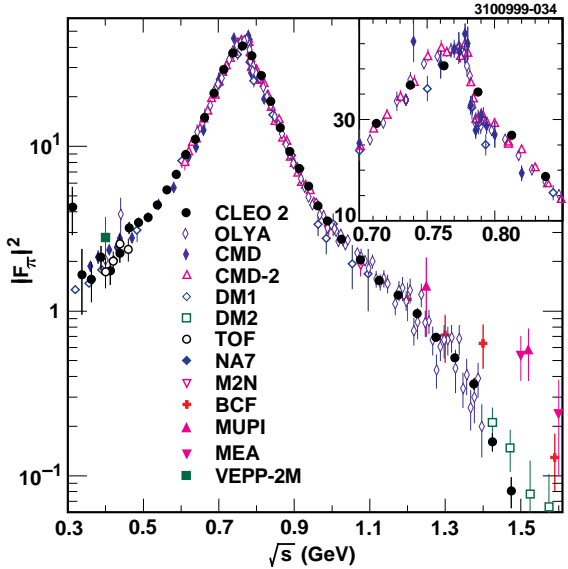}}}
\caption[]{\label{fig:fpi}
Comparison of $|F_\pi(q^2)|^2$ as determined from CLEO
$\tau$ data (filled circles), with that obtained from
$e^+e^-\to\pi^+\pi^-$ cross-sections measured
at VEPP and Adone.
}
\end{figure}

\begin{figure}[htb]
\centerline{\hbox{\epsfxsize75mm\leavevmode\epsfbox[94 322 498 714]
     {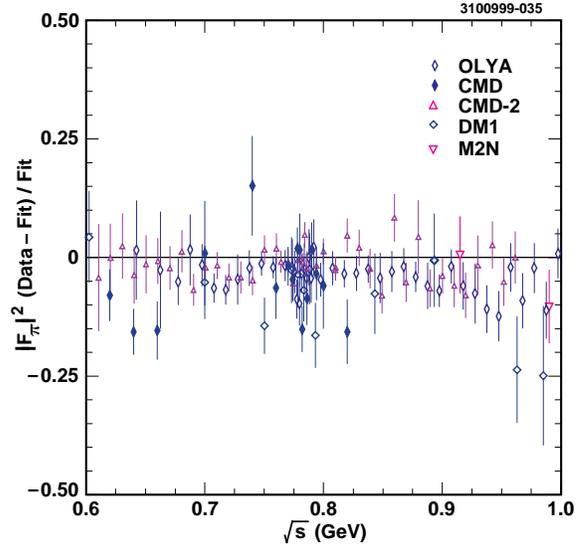}}}
\caption[]{\label{fig:dfpi}
Difference between
$|F_\pi(q^2)|^2$ as determined from CLEO
$\tau$ data, with that obtained from
$e^+e^-\to\pi^+\pi^-$ cross-section measurements.
}
\end{figure}

We see that the $\tau$ data follow the $e^+e^-$ data shape very well,
but the $e^+e^-$ data lie $\sim 3\%$ below $\tau$ data, throughout the spectrum.
This is also seen in the
fit of CLEO tau data to the G\&S model, where $|F_\pi(0)|^2 \simeq 1.03$,
and is consistent with the
discrepancy between the world average \cite{ref:pdg98} branching fraction
$\BR(\tau\to \pi\pi\nu_\tau)$ and the prediction
from CVC with the $e^+e^-$ data \cite{ref:eidelman98}:

\begin{eqnarray}
{
\BR(\tau\to\pi\pi^0\nu)} &=& (25.32\pm 0.15)\% \\
{ CVC:}               & & (24.52\pm 0.33)\% \\
{ \Delta}             &\sim& 3.2\pm 1.5\%.
\end{eqnarray}

If one assumes CVC, the data from tau decays on
$|F_\pi(q^2)|^2$ can be used to improve the prediction
for the contribution of hadronic vacuum polarization
to the anomalous magnetic moment of the muon,
$g_\mu-2$ \cite{ref:gm2}.
Such an improvement is much needed in order to 
make use of the precision of the 
BNL E821 $(g_\mu-2)$ experiment
to test for electroweak and beyond SM contributions.
The CLEO results can be used to improve this prediction,
but it has been suggested that allowing $|F_\pi(0)|$ to float
is not the appropriate way to make use of the data \cite{ref:davier}.

\section{$\tau\to 3\pi\nu_\tau$}

From conservation of G-parity and parity,
we expect the resonance $X$ in $\tau\to X\nu\to 3\pi\nu$ 
to have   $J^P = 0^-$ or $1^+$.
The vector ($J^P = 1^-$} current to $(3\pi)^-$ is forbidden by Bose symmetry
(there are two identical pions in the final state).
The axial-vector $1^+$ current is expected to be dominated by the
(broad and thus poorly understood) $a_1(1260)$ meson
($\Gamma_{a_1}\sim 400$~MeV).
There may also be a pseudoscalar ($0^-$)
current, \eg, $\pi^{\prime-} \to (3\pi)^-$;
but such a current is not conserved
( $P_\mu J^\mu_{0^-} \neq 0$),
so the $0^-$ current is suppressed by PCAC.

In the simplest models, \eg\ the K\"uhn \& Santamaria model \cite{ref:KandS},
the $a_1$ decays to $\rho \pi$ via S-wave.
From this, one predicts approximately equal rates to
$\pi^-\pi^+\pi^-$ and $\pi^-\pi^0\pi^0$.
K\"uhn and Wagner \cite{ref:KandW} pointed out that
the interference between amplitudes with identical pions
permits the measurement of the {\it sign} of the $a_1$ helicity,
and thus can be used to measure the parity-violating sign of the
$\tau^-\to\nu_\tau W^-$ coupling (the tau neutrino helicity $h_{\nu_\tau}$).

More sophisticated models, such as the 
Isgur, Morningstar, Reader \cite{ref:IMR} flux-tube-breaking
model, include D-wave $\rho\pi$, and $K^*K$ threshold effects;
these must be understood and taken into account in order to
accurately measure $h_{\nu_\tau}$.
They distort the mass-dependent width of the $a_1$ in measurable ways;
these must be understood and taken into account in order to
determine whether, \eg, a radially excited $a_1^\prime$ meson is present.
Other models, such as the Unitarized Quark Model of T\"ornqvist \cite{ref:torn},
suggests the possibility that scalar mesons participate in
the subsequent decay of the axial vector meson;
if observed, this could give some insight into the properties
of the (broad and thus poorly understood) scalar mesons.
Contributions from {\it iso}scalar mesons such as the tensor $f_2$
and scalar $f_0$ would also produce non-trivial relations between
the decay rates to $\pi^-\pi^+\pi^-$ and $\pi^-\pi^0\pi^0$.

Early work from ARGUS \cite{ref:ARGUS3pi} saw
significant { D-wave $\rho\pi$} production,
and hints of other contributions.
Delphi '97 \cite{ref:delphi3pi} saw anomalous substructure at high $M_{3\pi}$
--- a hint of { $a_1^\prime(1700)$}.
Fig.~\ref{fig:3piphenom} illustrates the potential complexity
of this decay.

\begin{figure}[htb]
{\footnotesize
\begin{center}
\setlength{\unitlength}{1.4cm}
\begin{picture}(5,2)
\put(0.2,1.6){\makebox(0,0){$\nu_\tau$}}
\put(0.2,0.4){\makebox(0,0){$\tau$}}
\put(0.2,1.0){\makebox(0,0){${\cal M}=$}}
\put(1.3,0.8){\makebox(0,0){$G_F /\sqrt{2}$}}
\put(2.0,1.2){\makebox(0,0)
             {$V_{ud} f_{a_1} \epsilon_\mu$}}
\put(2.5,0.6){\makebox(0,0) {  $BW(a_{1})$ }}
\put(2.45,0.3){\makebox(0,0){ { + $BW(a_{1}^\prime)$ (?)} }}
\put(2.45,0.0){\makebox(0,0){ { + $BW(\pi^\prime)$ (?)} }}
\put(2.5,-0.5){\makebox(0,0){$q= p_{\pi_1} - p_{\pi_{2/3}} \quad
                             Q= p_{\pi_1} + p_{\pi_{2/3}} - p_{\pi_{3/2}} \quad
                             P= p_{\pi_1} + p_{\pi_2} + p_{\pi_3}$  }}
\put(3.8,0.4){\makebox(0,0){$\pi$}}
\put(4.5,1.68){\makebox(0,0){$\pi$}}
\put(3.8,2.0){\makebox(0,0){$\pi$}}
\put(3.6,1.0){\makebox(0,0){$g_{a_1\rho\pi} \epsilon^\nu\epsbar_\nu$}}
\put(4.6,0.75){\makebox(0,0){{ +$\epsilon^\nu Q_\nu\epsbar_\mu P^\mu$ } }}
\put(4.8,0.50){\makebox(0,0){{ P-wave(?)} }}
\put(3.0,1.50){\makebox(0,0){$BW(\rho)$  }}
\put(4.3,1.50){\makebox(0,0){{ $+BW(f_2 )$ (?)}  }}
\put(4.3,1.25){\makebox(0,0){{ $+BW(f_0)$ (?)} }}
\put(4.4,1.88){\makebox(0,0)
             {$g_{\rho\pi\pi} \epsbar_\beta q^\beta$}}
\put(3.0,1.0){\line(1,-1){0.8}}   
\put(3.0,1.0){\line(1,1){0.7}}    
\put(3.68,1.68){\line(1,1){0.3}}    
\put(3.68,1.68){\line(1,0){0.6}}    
\put(1.0,1.0){\line(-1,1){0.8}}   
\put(1.0,1.0){\line(-1,-1){0.8}}  
\put(1.0,1.0){\line(1,0){0.6}}    
\put(1.6,1.0){\line(3,1){0.3}}    
\put(1.6,1.0){\line(3,-1){0.3}}   
\put(2.4,1.0){\oval(1.2,0.2)}     
\put(1.0,1.0){\circle*{0.05}}       
\put(1.6,1.0){\circle*{0.05}}       
\put(3.0,1.0){\circle*{0.05}}       
\put(3.68,1.68){\circle*{0.05}}       
\thicklines
\end{picture}
\end{center}
}
\caption[]{\label{fig:3piphenom}
Illustration of some of the many processes that can occur
in the decay $\tau^-\to 3\pi\nu_\tau$.
}
\end{figure}

\ssection{$\tau\to 3\pi\nu_\tau$ decay rate}

The decay rate can be described in terms of a matrix element squared
${\vert {\cal M} \vert }^2 $ given by:
\ben
\mbox{Lepton Tensor}\times \mbox{Hadron Tensor} = 
\een
\ben
L_{\mu\nu} \times J^\mu J^{\star\nu} =
       ( S_{\mu\nu} + i h_{\nu_\tau} A_{\mu\nu} )\times J^\mu J^{\star\nu} ,
\label{eqn:hnutau}
\een
where $S_{\mu\nu}$ is the symmetric part
and $A_{\mu\nu}$ the anti-symmetric part  of the lepton tensor,
fully known in the Standard Model.
The tau neutrino helicity is given by $h_{\nu_\tau} \equiv 
2 g_v g_a / (|g_v|^2+|g_a|^2)$ which is $-1$ in the standard $V-A$ model.
The hadronic current $J^\mu$ is a priori unknown,
but can be parameterized in a model-dependent way,
or in a model-independent way in terms of structure functions.

The Lorentz structure of $J_\mu$ is well-defined:
\begin{eqnarray}
J_\mu &=& 
 \left( -g_{\mu\nu} + \frac{ P_\mu P_\nu}{P^2} \right)
    \left[   ( p_{\pi_1} - p_{\pi_2} )^\nu F_1 \right. \nonumber \\
  &&  \left. + ( p_{\pi_1} - p_{\pi_3} )^\nu F_2
           + ( p_{\pi_2} - p_{\pi_3} )^\nu F_3 \right] \nonumber \\
  &&        + P_\mu F_4 
\end{eqnarray}

All the unknowns are in the form factors $F_i$,
which are modeled in terms of Breit-Wigner functions
for meson resonances, angular momentum factors (S,P,D...-wave),
and potentially finite meson radius effects or other effects.

The overall tau decay rate and $s \equiv m^2(3\pi)$ spectrum
is then given by
\begin{eqnarray}
d\Gamma(\nutau 3\pi) &=& \frac{G^2_F V^2_{ud}}{2 m_\tau} 
   \left[L^{\mu\nu} J_\mu J^*_\nu\right] dLips \\
 &=& 
\frac{G^2_F V^2_{ud}}{32\pi^2 m_\tau} (1 + 2 \frac{s}{m^2_\tau})
(1 - \frac{s}{m^2_\tau} ) \nonumber \\ 
& & \times \vert BW (s) \vert^2 \times
\frac{\Gamma_{3\pi} (s)}{s} ds,
\end{eqnarray}
where $BW(s)$ might be an overall Breit-Wigner line shape
for the $a_1$, and $\Gamma_{3\pi} (s)$ is the mass-dependent
decay rate to $3\pi$.

\ssection{$h_{\nu_\tau}$ from $\tau\rightarrow 3\pi\nu$}

K\"uhn and Wagner pointed out in 1984 \cite{ref:KandW}
that the parity-violating {\it signed} tau neutrino helicity $h_{\nu_\tau}$
can be measured using the decay $\tau\rightarrow 3\pi\nu$,
owing to its presence in Eqn.~\ref{eqn:hnutau}.
This requires an asymmetric part of the hadron tensor $J^\mu J^{\star\nu}$.
At least three pseudoscalars in final state are needed,
and an interference term between two amplitudes is needed.

There are two identical pions in this decay; thus, the 
$\rho$ can be formed in two ways:

\begin{equation}
\begin{array}{cllll}
\tau^- &\rightarrow &    a_1^-   \nu_\tau &\mbox{or}& a_1^-   \nu_\tau \\
                    & &  \,\,\hookrightarrow {\rho_1^0} \pi^-_2 
                    & &  \,\,\hookrightarrow {\rho_1^0} \pi^-_1 \\
                    & &  \quad\quad\hookrightarrow \pi^-_1  \pi^+ 
                    & &  \quad\quad\hookrightarrow \pi^-_2  \pi^+ 
\end{array}
\end{equation}

The imaginary part of the interference term between these two amplitudes
$\Im ( BW(\rho_1 ) \cdot BW ( \rho_2 )^\star )$ 
is a parity-odd observable that resolves 
the { left}- and { right}-handed part of the 
transverse polarization of the $a_1$:
\par
\setlength{\unitlength}{0.8cm}
\begin{picture}(9,1.2)(0,0)
\put(0.50,0.50){\vector(1,0){1.25}} 
\put(2.25,0.50){\vector(1,0){1.25}} 
\put(5.50,0.50){\vector(1,0){1.25}} 
\put(7.25,0.50){\vector(1,0){1.25}} 
\put(2.00,0.50){\circle*{0.15}} 
\put(7.00,0.50){\circle*{0.15}} 
\put(2.00,0.75){\makebox(0,0){$a_1$}}
\put(7.00,0.75){\makebox(0,0){$a_1$}}
\put(0.25,0.50){\makebox(0,0){$\tau$}}
\put(3.75,0.50){\makebox(0,0){$\nu_\tau$}}
\put(5.25,0.50){\makebox(0,0){$\tau$}}
\put(8.75,0.50){\makebox(0,0){$\nu_\tau$}}
\put(2.00,1.00){\makebox(0,0){{ $\DD\Longrightarrow$}}}
\put(7.00,1.00){\makebox(0,0){{ $\DD\Longleftarrow$}}}
\put(1.00,0.48){\makebox(0,0){{ $\DD\Rightarrow$}}}
\put(3.00,0.48){\makebox(0,0){{ $\DD\Leftarrow$}}}
\put(6.00,0.48){\makebox(0,0){{ $\DD\Leftarrow$}}}
\put(8.00,0.48){\makebox(0,0){{ $\DD\Rightarrow$}}}
\put(2.25,0.00){\makebox(0,0){{ left handed $\nu_\tau$}}}
\put(7.25,0.00){\makebox(0,0){{ right handed $\nu_\tau$}}}
\end{picture}
\par
A measurement of this parity-odd term in the decay rate permits
a measurement of $h_{\nu_\tau}$, so long as the parity-even
(dominant) part of the decay rate is well and truly modeled.

\ssection{CLEO results on $\tau\to 3\pi\nu_\tau$}

CLEO has recently published two papers on our   analysis of
the $\tau^-\to \nu_\tau \pi^-\pi^0\pi^0$ decay.
This decay is favored over the all-charged 
$\nu_\tau\pi^-\pi^+\pi^-$ decay, despite lower statistics,
because of less background (from $K\pi\pi$, $KK\pi$, $4\pi$, and
hadronic events), and because isoscalar decays such as 
$f_2\to\pi\pi$ have one entry per event in $a_1\to \pi^-f_2\to \pi^-\pi^0\pi^0$
but two in $\pi^-\pi^+\pi^-$.

In both papers, the full CLEO II sample of $\approx 4.3\times 10^6$ $\tau^+\tau^-$ pairs
was used, resulting in
$30800$ $\tau^\mp\rightarrow\pi^\mp\pi^0\pi^0\nu$ events (all tags), and
$14600$ $\tau^\mp\rightarrow\pi^\mp\pi^0\pi^0\nu$ lepton tagged events.
The background is $\approx 10\%$, mostly $\tau\to 4\pi \nu$ and fake $\pi^0$'s.

The first paper \cite{ref:cleo3pimd}
describes a model dependent analysis,
in which the Dalitz plot distribution ($s_1 \equiv m^2(\pi^-\pi^0_1)$ 
vs $s_2 \equiv m^2(\pi^-\pi^0_2)$) is fitted, in bins of
$s \equiv m^2(\pi^-\pi^0_1\pi^0_2)$, for contributions from
different intermediate resonances and angular momenta.
All measurable observables of production
are also included in these fits.
This results in a measured total width
$\Gamma_{3\pi} (s)=\int J_\mu J^{\star\mu} ds_1 ds_2$.
There is clear evidence for contributions from isoscalar
intermediate resonances: $\sigma\pi$, $f_0\pi$, $f_2\pi$.
There is also clear evidence of $K^*K$ threshold
turn-on in the $\Gamma_{3\pi}(s)$ mass dependence.
In a second step, the overall lineshape $BW(s)$ is 
determined by measuring the invariant mass distribution
of the three pions.
In this paper, a measurement of the parity-violating signed 
neutrino helicity $h_{\nu_\tau}$, and 
model-dependent limits on scalar and vector $3\pi$ currents,
are presented.

\begin{figure}[htb]
\centerline{
\hbox{\epsfxsize75mm\leavevmode\epsfbox[0 318 603 497]
{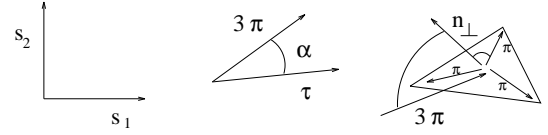}}
}
\caption[]{\label{fig:angles}
Kinematical variables characterizing the $3\pi$ system
in $\tau\to 3\pi\nu_\tau$ decay:
The Dalitz plot variables, and
two angles that are measurable in the decay 
which are sensitive to its Lorentz structure.
}
\end{figure}

The second paper \cite{ref:cleo3pimi}
describes a model independent analysis,
in which the squared matrix element ${\vert {\cal M} \vert}^2$
is parameterized in terms of a sum of 16 independent terms \cite{ref:strucfun}:
\ben
\vert {\cal M} \vert^2 = L_{\mu\nu} \times J^\mu J^{\star\nu} =
       \sum_{X=1}^{16} L_X W_X ,
\een
where the $L_X$ are 16 well-defined functions of decay observables,
designed to select contributions to the overall hadronic current
with different Lorentz structure (axial-vector, vector, scalar, \etc),
and the $W_X$ are 16 structure functions, functions of
$s$, $s_1$, and $s_2$, that parameterize the hadronic dynamics.
In this paper, model-independent limits on scalar and vector $3\pi$ currents
are presented.

\ssection{Model dependent analysis}

In this analysis, the substructure in the hadronic current $J^\mu$
is determined in the context of a model, via a
Likelihood fit to the Dalitz plot in full kinematical space,
in bins of $m_{3\pi}$. 
The variables used are 
$s=m^2(3\pi)$, $s_1 = m^2(\pi^-\pi^0_1)$, $s_2 = m^2(\pi^-\pi^0_2)$;
and the angular observables 
$\alpha$, $\psi$ defined in Fig.~\ref{fig:angles}.

The amplitudes used in the fit to the $3\pi$ substructure
which were found to be significant were:
\begin{itemize}
\item $J_1^\mu$: s-wave $1^+ \to \rho\pi$
\item $J_2^\mu$: s-wave $1^+ \to \rho^\prime\pi$ 
\item $J_3^\mu$: d-wave $1^+ \to \rho\pi$ 
\item $J_4^\mu$: d-wave $1^+ \to \rho^\prime\pi$
\item $J_5^\mu$: p-wave $1^+ \to f_2(1275)\pi$ 
\item $J_6^\mu$: p-wave $1^+ \to f_0 (400-1200) \pi$, \\
denoted as $\sigma\pi$ 
\item { $J_7^\mu$}: p-wave amplitude of $1^+ \to f_0(1370)\pi$ 
\end{itemize}

The mass and width of the $f_0 (1370)$ and $ f_0 (400-1200)$ ($\sigma$) 
were fixed according to T\"ornqvist's Unitarized Quark Model \cite{ref:torn}:
\begin{center}
$m_{f_0 (1370)} = 1.186\mbox{ GeV/c}^2$; \\
$\quad\Gamma_{f_0 (1370)} = 0.350\mbox{ GeV}$; \\
$m_\sigma       = 0.860\mbox{ GeV/c}^2$; \\
$\quad\Gamma_\sigma       = 0.880\mbox{ GeV}$ .
\end{center}

The total current was parameterized as a coherent sum of these 
contributions:
\ben
J^\mu = \sum_{i=1}^{i=7}  \beta_i \times J^\mu_i \times F_i,
\een
where the $\beta_i$ are (complex) fit parameters,
and the $F_i$ are form factors to take into account the finite size
of the mesons involved:
$F_i = e^{-0.5 R^2 p^{\star 2}_i}$.
In the nominal fit, $R$ was fixed at $0$ (so that $F_i = 1$);
in other fits, $R$ was allowed to vary.

\begin{figure}[htb]
\begin{center}
\leavevmode
\unitlength1.0cm
\epsfxsize=75mm
\epsffile{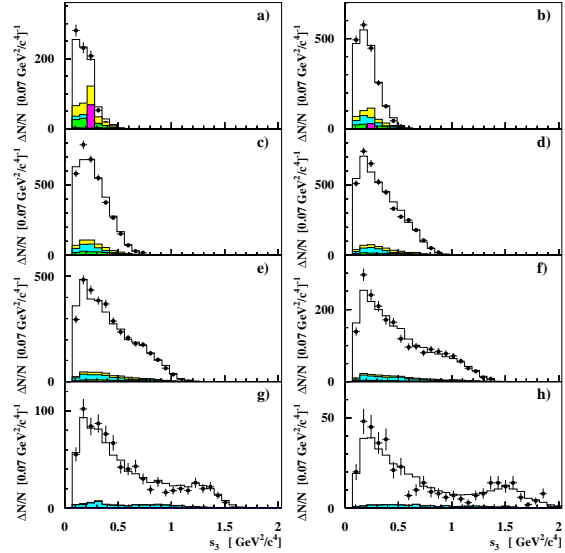}
\end{center}
\caption[]{\label{fig:3pis3}
Results of the fit to the $3\pi$ substructure in 
$\tau^\mp\to \pi^\mp\pi^0_1\pi^0_2\nu_\tau$ decays.
Shown are projections of the data and the fit in 
$s_3=m^2_{\pi^0_1 + \pi^0_2} $.
In each plot, the points with error bars are the data,
and the histograms are the results of the fit.
The plots are for different values of $m_{3\pi}$:
(a) $0.6-0.9$ GeV;
(b) $0.9-1.0$;
(c) $1.0-1.1$;
(d) $1.1-1.2$;
(e) $1.2-1.3$;
(f) $1.3-1.4$;
(g) $1.4-1.5$;
(h) $1.5-1.8$.
}
\end{figure}

\begin{figure}[htb]
\begin{center}
\leavevmode
\unitlength1.0cm
\epsfxsize=75mm
\epsffile{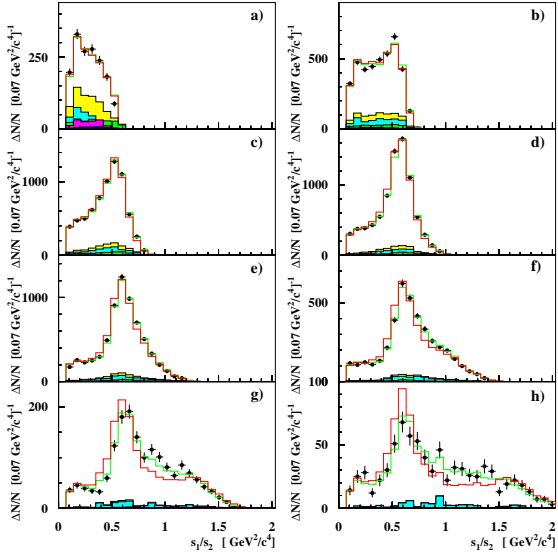}
\end{center}
\caption[]{\label{fig:3pis1}
Results of the fit to the $3\pi$ substructure in 
$\tau^\mp\to \pi^\mp\pi^0_1\pi^0_2\nu_\tau$ decays.
Shown are projections of the data and the fit in 
$s_1=m^2_{\pi^-   + \pi^0_1} $ and   
$s_2=m^2_{\pi^-   + \pi^0_2} $ (two entries per event).
In each plot, the points with error bars are the data,
filled histograms are background contributions,
and the clear histograms are the results of the fit.
The plots are for different values of $m_{3\pi}$,
as in Fig.~\protect\ref{fig:3pis3}.
}
\end{figure}

The results of these fits are illustrated in Fig.~\ref{fig:3pis3}
and \ref{fig:3pis1}.
Good fits ($<3\sigma$) are obtained in all $m_{3\pi}$ bins.
There is clear evidence in the $s_3$ projections at high $s$
for the isoscalar $f_2(1275)$.
The fit results for the various contributions to the total 
$\tau^-\to \pi^-\pi^0\pi^0\nu_\tau$ decay rate
are given in Table~\ref{tab:3pisub}.

\begin{table}
\caption[]{\label{tab:3pisub}
Contributions to the total
$\tau^-\to \pi^-\pi^0\pi^0\nu_\tau$ decay rate,
and their significance.
}
\begin{center}
\renewcommand{\arraystretch}{0.8}
\tabcolsep 2pt
{
\begin{tabular}{l l  c c}
             &        &    Significance    & $\mbox{${\cal B}$ fraction}$$(\%)$ \\ \hline 
$\rho      $ & s-wave &                    & $ 69.4$                 \\
$\rho(1370)$ & s-wave & $1.4\sigma$        & $ 0.30\pm 0.64\pm 0.17$ \\
$\rho      $ & d-wave & $5.0\sigma$        & $ 0.36\pm 0.17\pm 0.06$ \\
$\rho(1370)$ & d-wave & $3.1\sigma$        & $ 0.43\pm 0.28\pm 0.06$ \\
$f_2 (1275)$ & p-wave & $4.2\sigma$        & $ 0.14\pm 0.06\pm 0.02$ \\
$\sigma    $ & p-wave & ${ 8.2\sigma}$ & ${ 16.18\pm 3.85\pm 1.28}$ \\
$f_0 (1186)$ & p-wave & $5.4\sigma$        & $ 4.29\pm 2.29\pm 0.73$ 
\end{tabular}
}
\end{center}
\end{table}

We note that the $\rho\pi$ s-wave with ${\cal B}\approx 70\%$ is
dominant, as expected.
With the exception of $\rho^\prime\pi$ s-wave, all amplitudes are significant;
the K\&S model is too simple.
Isoscalars contribute 
to the $3\pi$ hadronic current
with { ${\cal B}\approx 20\%$};
especially the $\sigma$ cannot be neglected.
Curiously, the $\rho^\prime$ shows up more strongly in d-wave than in s-wave.
Finally,
the couplings to each of the $J_i$ sub-currents are consistent with being
constant as a function of $m_{3\pi}$,
suggesting that only one resonance (the $a_1$) is responsible 
for all this substructure.

There is no evidence for scalar $3\pi$ currents,
and the following model-dependent upper limits at $90\%$ CL
are derived:
\ben
{\cal B} ( \tau\to\pi^\prime\nu\to\rho\pi\nu\to 3\pi\nu) <
1.0\times 10^{-4} 
\een
\ben
{\cal B} ( \tau\to\pi^\prime\nu\to\sigma\pi\nu\to 3\pi\nu) <
1.9\times 10^{-4} 
\een
\ben
h_{\nu_\tau} = -1.02\pm 0.13\pm 0.03  \;\mbox{(SM = $-$1)}
\een

These fit results can be converted to a model for the 
all charged mode, $\tau^\mp\to\pi^\mp\pi^\mp\pi^\pm\nu$.
The conversion is non-trivial, 
due to the presence of isoscalars with different isospin structure:
\begin{eqnarray}
|0,0\rangle =  \frac{1}{\sqrt{3}} |1,+1\rangle |1,-1\rangle && \nonumber \\
             +\frac{1}{\sqrt{3}} |1,-1\rangle |1,+1\rangle 
          - \frac{1}{\sqrt{3}} |1, 0\rangle |1,0 \rangle . &&
\end{eqnarray}

Although this fit procedure was not applied to the 
$\approx 80000$ $\tau^\mp\to\pi^\mp\pi^\mp\pi^\pm\nu$ events
selected from the CLEO data set,
qualitative agreement with the Dalitz plot distributions
is excellent. In addition, the total decay rate 
predicted for $\tau^\mp\to\pi^\mp\pi^\mp\pi^\pm\nu$ 
is in excellent agreement with the world average measured values \cite{ref:pdg98}.

To model the full
three pion mass spectrum,
we form a coherent superposition of
contributions from the $a_1$ and a radially excited $a_1^\prime$:
\begin{eqnarray}
B(s)&=& B_{a_1} (s) + \kappa\cdot B_{a^\prime_1} (s) \\
&=& \frac{1}{s-m^2_{a_1}(s)+im_{0\,a_1}\Gamma^{a_1}_{tot} (s)} \nonumber\\  
&+&\frac{\kappa}{s-m^2_{0\, a_1^\prime} +im_{0\,a_1^\prime} \, 
\Gamma^{a_1^\prime}_{tot}(s)}.
\label{eqn:a1prime}
\end{eqnarray}
The total width $\Gamma_{tot} (s)$ must be modeled carefully.
We integrate the fit results for $J^\mu$ over the Dalitz plot
in bins of $m(3\pi)$, and include contributions to the total width
from $K^\star K\to KK\pi$ and $f_0 (980)\pi\to KK\pi$:
\begin{eqnarray}
\Gamma_{tot} (s) &=& \Gamma_{2\pi^0\pi^-} (s) + 
                   \Gamma_{2\pi^-\pi^+} (s)  \nonumber \\
                 &&  +\Gamma_{K^\star K} (s)   + 
                   \Gamma_{f_0 (980)\pi} (s)  
\end{eqnarray}

If the total width runs (is a function of $s$), then the 
mass $m^2(s)$ can run as well:
\ben
m^2(s) = m_0^2 + \frac{1}{\pi} \int_{s_{th}}^\infty 
\frac{ m_0 \Gamma_{tot} (s\prime )}{(s- s\prime)} ds\prime  
\een

Thus we perform $\chi^2$ fits to the total $m(3\pi)$ spectrum
with and without $a_1^\prime$;
with $KK\pi$ contributions to $\Gamma_{tot}$ or not; 
with running mass or constant mass; 
and with or without finite meson radii.

\begin{figure}[htb]
\centerline{\psfig{figure=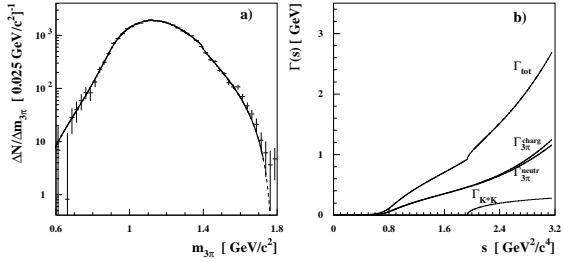,width=75mm}}
\caption[]{\label{fig:a1fit}
(a) Nominal fit to the $m(3\pi)$ spectrum
with no $a_1^\prime$.
Points with errors are the data; solid line is the fit result.
Note the kink around $m_{3\pi}=1.4$ GeV
due to the turn-on of the $K^\star K$ threshold
in the total width.
(b) Calculated contributions to the total width $\Gamma_{3\pi}(s)$.
}
\end{figure}

We obtain good fits with either 
constant or running mass, with and without $f_0 (980)\pi$,
and for meson radii anywhere in the range $0\le R \le 2\mbox{ GeV}^{-1}$.
The $K^\star K$ threshold is needed for good fit.
The best fits are obtained  for $R$ in the range
$1.2\le R \le 1.4\mbox{ GeV}^{-1}$.
Our nominal fit is chosen as the one with
constant $a_1$ mass, $K^\star K$ threshold included, 
no $f_0 (980) \pi$ threshold, and $R=0$.
The fit results are shown in Fig.~\ref{fig:a1fit}.
We obtain
$m_{a_1} = 1.331\pm 0.010\pm 0.003$,
$\Gamma_{a_1} = 0.814\pm 0.036\pm 0.013$,
${\cal B} ( a_1 \to K^\star K ) = (3.3\pm 0.5\pm 0.1)\%$,
with a $ \chi^2 =  39.3 / 41 \mbox{ dof} $.
Here, again, we observe consistent results with 
the all-charged mode $\tau^-\to \pi^-\pi^+\pi^-\nu_\tau$.

It is important to note that the parameters of the $a_1$
resulting from this fit have a large model dependence,
which is not included in the systematic errors.
These parameters will change if additional, higher-mass contributions
are added to the fit.

\begin{figure}[htb]
\centerline{\psfig{figure=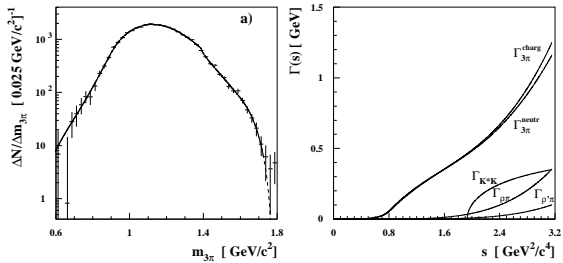,width=75mm}}
\caption[]{\label{fig:a1pfit}
Fit to the $m(3\pi)$ spectrum
including the $a_1^\prime$.
See caption to Fig.~\protect\ref{fig:a1fit}.
}
\end{figure}

There appears to be a small excess of data at high $s$,
suggesting the presence of an $a_1^\prime$
and/or more thresholds in the total width.
We have therefore performed a fit including the $a_1^\prime$,
as in Eqn.~\ref{eqn:a1prime}.
We are not sensitive to the mass or width of the $a_1^\prime$,
and thus fix them to be
$m_{a_1^\prime} = 1700$~MeV, $\Gamma_{a_1^\prime} = 300$~MeV.
The fit results are shown in Fig.~\ref{fig:a1pfit}.
We obtain ${ |\kappa| = 0.053 \pm 0.019}\ \mbox{ (stat.)}$ 
with phase $\phi_\kappa$ consistent with zero, and 
{ $\chi^2$  $=  28.9/39$}~dof.
The improvement of fit yields a
significance of $2.9\sigma$ for the $a_1^\prime$, and
${\cal B} ( \tau\to a_1^\prime\nu ) = (1.6\pm 1.1\pm 0.3\pm 0.7)\times 10^{-4}$.
More statistics needed to conclusively state if the $a_1^\prime$
participates or not in $\tau\to 3\pi\nu$ decay.

\ssection{$\tau\to 3\pi\nu_\tau$ Structure functions}

CLEO has analyzed its $14600$ $\tau^\mp\to\pi^\mp\pi^0\pi^0\nu$ lepton tagged events
in terms of the structure functions defined in Ref.~\cite{ref:strucfun}.
The decay rate is written as:
\ben
d\Gamma_{3\pi\nu_\tau}  \propto  
  \sum_X^{16} \bar{L}_X (\alpha ,\beta ,\gamma ) W_X (s, s_1 , s_2 ) d\mbox{Phs}
\een
\ben
\mbox{ with } X \in \{A,B,\ldots, I,SA,SB,\ldots , SG \}.
\een
The 16 structure functions $W_X$ that contain all the
information on the hadronic structure 
depend on $s$, $s_1$, and $s_2$ only. 
We can measure these structure functions independent of any model.
They can be interpreted by comparing them with predictions
from a model for $J^\mu$, such as the one resulting from the 
fit described above.

In the $3\pi$ rest frame, with $z$-axis perpendicular
to the $3\pi$ decay frame,
the hadronic current $h^\mu$ has a time-like component
$h^0$ from pseudoscalar currents (such as the $\pi^\prime$),
a component along the $z$ axis, $h^3$, from vector currents
(such as $\rho^\prime\to\rho\pi$ via the Wess-Zumino anomaly),
and transverse components $h^1$ and $h^2$ from the dominant
axial-vector current (such as the $a_1$).
The leptonic tensor components $L_X$ are defined so as to 
decompose the hadronic current into contributions
from the different $J^P$-states (scalar, vector, axial)
according to Table~\ref{tab:struc}.
The measurement of the structure functions
$W_{SA}$ and $W_B$  allows us to determine the non-axial
vector contributions model independently.
More directly, we can look for non-zero contributions
to the real or imaginary parts of $h^0$ or $h^3$
in bins of $s$, $s_1$, and $s_2$.

\begin{table}
\caption[]{\label{tab:struc}
Composition of the structure functions in terms of the 
spin and parity $J^P$ of the hadronic current.
}
{\footnotesize
\begin{center}
\begin{tabular}{ c  c  c   c }
  & $J=0$ & $J^P = 1^+$  & $J^P = 1^-$ \\ 
  & $h^0$ & $h^1$, $h^2$ & $h^3$       \\ \hline
{$J=0$}           & ${\bf W_{SA}}$ &  & \\ 
{$h^{\star 0}$}  \\ \hline
{$J^P = 1^+$}  & 
{$W_{SB}$, $W_{SC}$}  &
{${\bf W_A}$} & \\
{$h^{\star 1} $, $h^{\star 2} $}  &  
{$W_{SD}$, $W_{SE}$} & 
{$W_C$, $W_D$, $W_E$} & \\ \hline
{$J^P = 1^-$}  & $W_{SF}$, $W_{SG}$ &
{$W_F$, $W_G$} & ${\bf W_B}$ \\
{ $h^{\star 3}$}  & & 
{$W_H$, $W_I$}   & \\
\end{tabular}
\end{center}
}
\end{table}

\begin{figure}[htb]
\centerline{\psfig{figure=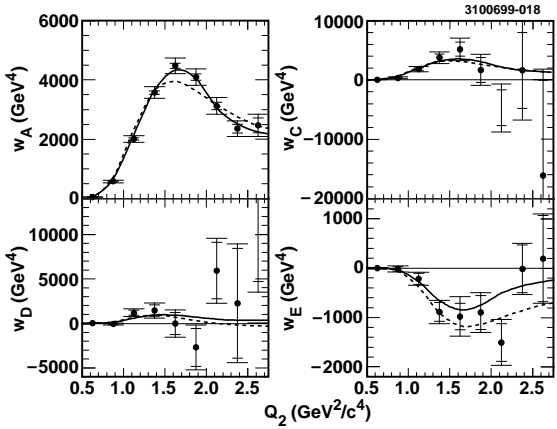,width=75mm}}
\caption[]{\label{fig:cleostruca}
The structure functions $W_A(s)$, $W_C(s)$, $W_D(s)$, and $W_E(s)$
integrated over $s_1$ and $s_2$, as measured by CLEO.
The dashed line is the prediction from the K\&S model,
and the solid line is the prediction from the CLEO
model-dependent fit.
}
\end{figure}

The structure functions $W_A(s)$, $W_C(s)$, $W_D(s)$, and $W_E(s)$
remain non-zero when one integrates over the Dalitz plot variables
$s_1$ and $s_2$. Their measured distribution is shown in 
Fig.~\ref{fig:cleostruca}.
The data are compared with the K\&S model and with the 
results of the CLEO fit to the more elaborate
model described above.

\begin{figure}[htb]
\centerline{\psfig{figure=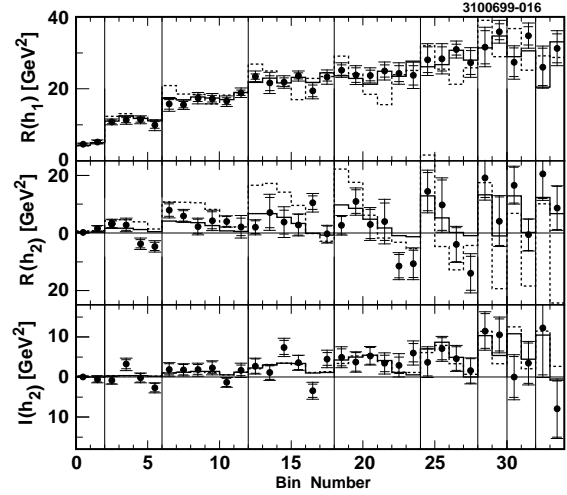,width=75mm}}
\caption[]{\label{fig:cleostrucb}
The real and imaginary parts of the $3\pi$ hadron current 
induced by axial vector current, as measured by CLEO.
Dashed line is the prediction from the K\&S model,
and the solid line is the prediction from the CLEO
model-dependent fit.
}
\end{figure}

\begin{figure}[htb]
\centerline{\psfig{figure=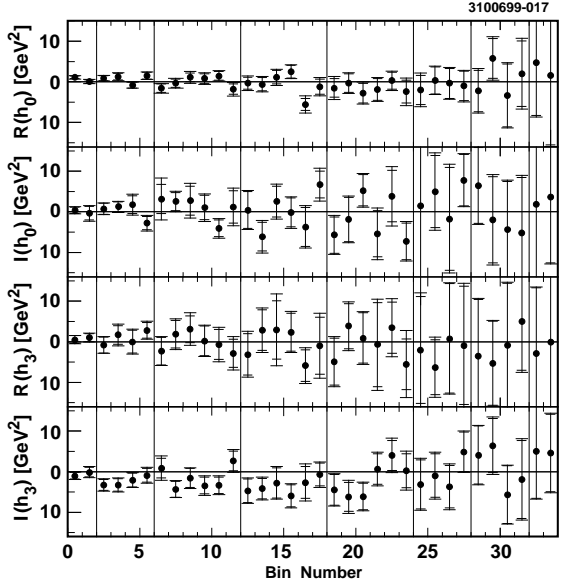,width=75mm}}
\caption[]{\label{fig:cleostrucc}
The real and imaginary parts of the $3\pi$ hadron current 
induced by the scalar ($h^0$) and vector ($h^3$) currents, as measured by CLEO.
}
\end{figure}

The components of the hadronic current $h^\mu$
as described above can be measured in the full
$s$, $s_1$, and $s_2$ space. Choosing a set of bins
in this 3D space as described in Ref.~\cite{ref:cleo3pimi},
we obtain the results shown in Figs.~\ref{fig:cleostrucb}
and \ref{fig:cleostrucc}. The axial-vector induced components
of the hadronic current are well described by the CLEO 
model-dependent fit, while the non-axial-vector components
are consistent with zero everywhere.
From this, we extract a model-independent limits on
scalar and vector contributions to the $\tau^-\to 3\pi\nu_\tau$ decay,
at 95\%\ CL:
{\small 
\ben
{\cal B} ( \tau^\mp \to S \nu \to (3\pi)^\mp \nu )
       /{\cal B} ( \tau^\mp \to (3\pi)^\mp\nu ) < 9.4\% 
\een
\ben
{\cal B} ( \tau^\mp \to V \nu \to (3\pi)^\mp \nu )
      /{\cal B} ( \tau^\mp \to (3\pi)^\mp \nu ) < 7.3\%
\een
}

From the measured hadronic current, 
all sixteen structure functions in the $\tau^-\to 3\pi\nu_\tau$ decay
have been determined, for the first time.

\ssection{Summary on $\tau\to 3\pi\nu_\tau$}

The high statistics CLEO analyses of $\tau\to 3\pi\nu_\tau$
are permitting detailed studies of the hadronic substructure,
precision measurements of signed $\nu_\tau$ helicity,
and have revealed significant contributions 
to the $3\pi$ system other than $a_1\to \rho \pi$.
The model-dependent fits to full kinematical distribution reveal
significant signals for isoscalars ($f_0$, $f_2$, and $\sigma$),
clear  evidence for $K^\star K$ threshold,
weak evidence for $a_1^\prime$,
and limits on the PCAC-violating $\pi^\prime$.
The model-independent structure function analyses 
give limits on non axial vector contributions,
and clean tests of the models.
Still, there are many open questions:
Can we learn more about the 
$a_1$ lineshape: running/constant mass, thresholds, \etc?
Is there an $a_1^\prime$? How does it decay?
Are there other components to the substructure?
A detailed analysis of the higher-statistics 
all-charged $\tau^-\to\pi^-\pi^+\pi^-\nu_\tau$ mode may 
shed more light on these questions.

\section{\bmath $\tau^-\to (4\pi)^- \nu_\tau$ }

We expect the $\tau\to 4\pi\nu_\tau$ decay to proceed via
the vector ($J^P = 1^-$) current,
dominated by the $\rho$ meson and its radial excitations
$\rho'$, $\rho''$, \etc.
Axial vector currents are ``second-class'' (isospin violating);
an example is $\tau\to b_1\nu_\tau $, $b_1\to \omega\pi$.
Given the large phase space for the $4\pi$ system,
even the simplest models are already complicated!

CLEO has analyzed these decays with the goals of
extracting the parameters (mass and pole width) of the $\rho'$;
searching for second class (axial) currents;
exploring the resonant decomposition of $4\pi$ system 
($\omega\pi$, $\eta\pi$, $a_1\pi$);
and testing CVC by comparing the cross sections for
$e^+e^-\to 2\pi^+2\pi^-$, $\pi^+\pi^-2\pi^0$
at low energy 
to { $\tau^- \to  \nu_\tau 2\pi^-\pi^+\pi^0$, $\pi^-3\pi^0$.

A good model of the $4\pi$ spectral function
is needed in order to reliably extract a limit on 
the tau neutrino mass from the $4\pi$ kinematical distributions.
CLEO \cite{ref:mnu4pi} set the limit
$m_{\nu_\tau} < 28$ MeV/c$^2$}, 95\%\ CL;
the  4 MeV model-dependence 
dominates the systematic error in that measurement.

The decay $\tau\to\omega\pi\nu$ first measured by 
ARGUS \cite{ref:argusomegapi} and CLEO \cite{ref:cleoomegapi} in 1987.
The $\rho\pi\pi$ branching fractions were measured by
ARGUS \cite{ref:argusrhopipi} in 1991, and ALEPH in 1997 \cite{ref:alephrhopipi}.
The mass and width of $\rho'$ were extracted from $\tau\to\pi\pi^0\nu$ 
by ALEPH in 1997 \cite{ref:aleph2pi} and CLEO in 1999 \cite{ref:cleorho}.

\ssection{$\tau\to 4\pi\nu_\tau$ from CLEO} 

\begin{figure}[htb]
\centerline{\psfig{figure=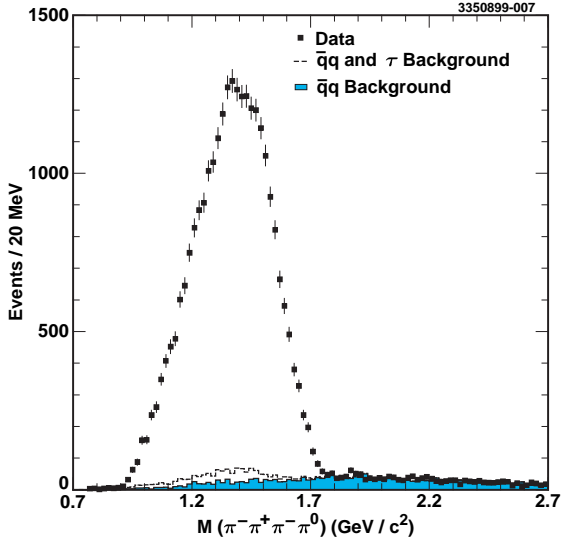,width=75mm}}
\caption[]{\label{fig:cleom4pi}
The $m(4\pi)$ mass spectrum in $\tau^\mp\to \pi^\mp\pi^\pm\pi^\mp\pi^0\nu_\tau$
as measured by CLEO, before background subtraction.
}
\end{figure}

The CLEO analysis \cite{ref:cleo4pi} uses the 
CLEO-II data set of $N_{\tau\tau} \approx 4.3\times 10^6$
produced tau pairs, and selects
$\sim 24,000$ events consistent with $\tau^\mp\to \pi^\mp\pi^\pm\pi^\mp\pi^0\nu_\tau$.
The $m(4\pi)$ mass spectrum is shown in Fig.~\ref{fig:cleom4pi}.
After subtracting estimated backgrounds dominated by
$K\pi\pi\pi\nu$, $KK\pi\pi\nu$, $K_S\pi\pi\nu$,
we extract the branching fraction
\ben
\BR(\tau^-\to 3\pi\pi^0\nu_\tau) = (4.19\pm0.10\pm0.21)\%.
\een

\ssection{\bmath $\omega\pi$ Spectral Function }

We construct the spectral function { $V^{3\pi\pi^0}(q)$, $q=m(3\pi\pi^0)$}
by correcting for background, efficiency, and
the production dynamics:
\begin{eqnarray}
\begin{array}{lr}
V^{3\pi\pi^0}(q) = &\frac{1}{N} \frac{dN(q)}{dq} 
\frac{1}{q(M_{\tau}^2-q^2)^2(M_{\tau}^2+2q^2)} \\
& \times \frac{\BR(\tau\to 3\pi\pi^0\nu_{\tau})}{
\BR(\tau\to e\bar{\nu_{e}}\nu_{\tau})}
\frac{M_{\tau}^8}{12\pi V_{ud}^2}.
\end{array}
\end{eqnarray}

\begin{figure}[htb]
\centerline{\psfig{figure=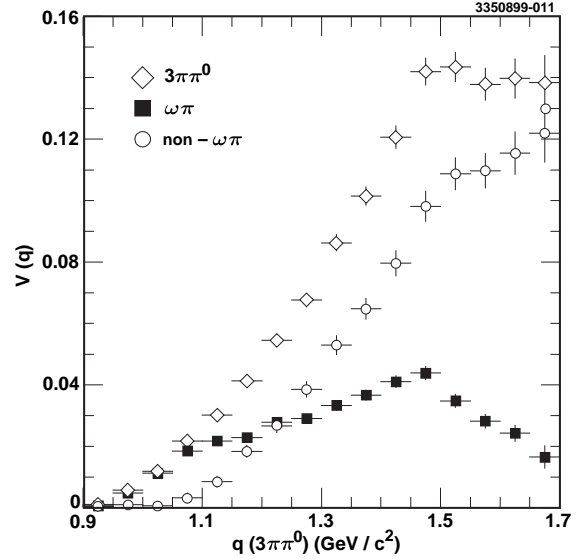,width=75mm}}
\caption[]{\label{fig:cleos4pi}
The spectral functions in bins of $q=M(3\pi\pi^0)$
as measured by CLEO:
the total $V^{3\pi\pi^0}(q)$ (open diamonds),
$V^{\omega\pi}(q) $ (filled squares),
and $V^{non-\omega\pi}(q)$ (open circles).
}
\end{figure}

Focusing on $\tau^\mp\to \omega\pi^\mp\nu_\tau$,
we extract the $\omega$ signal in bins of $q=M(3\pi\pi^0)$,
and form the $V^{\omega\pi}(q) $ spectral function, and the remainder:
\ben
V^{non-\omega\pi}(q) \equiv V^{3\pi\pi^0}(q)-V^{\omega\pi}(q).
\een
These are shown in Fig.~\ref{fig:cleos4pi}.

\begin{figure}[htb]
\centerline{\psfig{figure=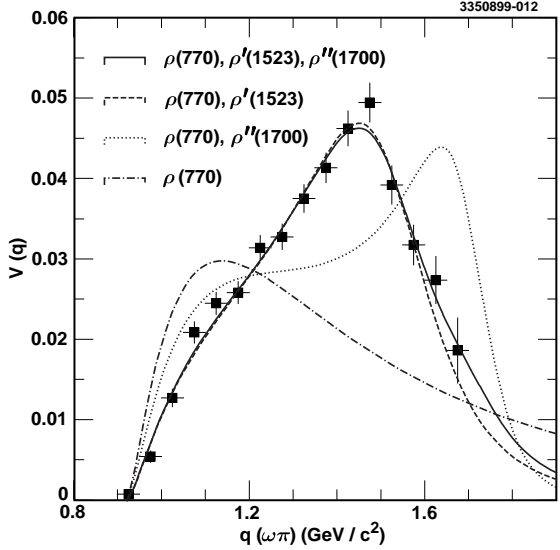,width=75mm}}
\caption[]{\label{fig:cleoopifit}
Fits to the $V^{\omega\pi}(q) $ spectral function
measured by CLEO in $\tau\to\omega\pi\nu_\tau$.
}
\end{figure}

We fit the $\omega\pi$ spectral function using a coherent sum of
combinations of 
Breit-Wigner lineshapes with mass dependent widths, for 
$\rho(770)$, $\rho'$, and $\rho^{\prime\prime}(1700)$.
We take the parameters of the $\rho$ and $\rho''$ as known:
\begin{eqnarray}
M_{\rho}=770\ \mbox{MeV}/c^2\ ;    & \Gamma_{\rho}=151\ \mbox{MeV}/c^2\ ;  \\
M_{\rho''}=1700\ \mbox{MeV}/c^2\ ; & \Gamma_{\rho''}=235\ \mbox{MeV}/c^2\ .
\end{eqnarray}
The results of these fits are shown in Fig.~\ref{fig:cleoopifit}.

Good fits are obtained only when both the { $\rho(770)$} and { $\rho(1450)$}
are included. Results with the $\rho^{\prime\prime}(1700)$
included or not included give consistent results for the 
$\rho'$ parameters:
\ben
M_{\rho'}=(1.523\pm0.010)\ \mbox{GeV}/c^2  
\een
\ben
\Gamma_{\rho'}=(0.400\pm0.035)\ \mbox{GeV}/c^2.
\een

Recall from the $2\pi$ channel, we had $M_{\rho^\prime} \sim 1400$~MeV.
The PDG values: are
\ben
M_{\rho'}=(1.465\pm0.025)\ \mbox{GeV}/c^2,
\een
\ben
\Gamma_{\rho'}=(0.310\pm0.060)\ \mbox{GeV}/c^2,
\een
dominated by proton experiments on fixed target, $e^+e^-\to\pi^+\pi^-$ and
$e^+e^-\to\eta\pi^+\pi^-$, as well as from earlier $\tau\to\pi\pi^0\nu$ results.
The origin of these differences is unresolved.

\ssection{\bmath Second Class Currents in $\tau\to\omega\pi\nu$}

There are two { axial-vector ($J^P = 1^+$)} states: 
The $a_1(1260)$ in the $^3P_1$ octet, with $J^{PG} = 1^{+-}$,
couples to the $W$ as a ``first-class'' current;
the $b_1(1235)$ in the $^1P_1$ octet, with $J^{PG} = 1^{++}$,
doesn't couple to the $W$ (``second-class'' current)
except via isospin (G-parity) violation (the weak decay constant
$f_{b_1}\approx 0$).

The $a_1$ decays to  $\rho\pi$ via S-wave, thence to $3\pi$.
The $\rho^\prime$ decays to $\omega\pi$  via P-wave, thence to $4\pi$.
The $b_1$ decays to $\omega\pi$ via S-wave, thence to $4\pi$.
The difference in G-parity for the states which decay to $4\pi$
is reflected in the different expected polarization
of the $\omega$ meson, and thus in the angular distribution of
the angle between the normal to the $\omega$ decay plane
and the direction of the 4th pion (``helicity angle'') 
$\cos\chi = \hat{n}_\perp^\omega \cdot \hat{p}_{\pi_4}$.
The different expected angular distributions are given in
Table~\ref{tab:chi4pi}.
   
\begin{table}[htb]
\caption[]{\label{tab:chi4pi}
Expected distributions of the $\omega$ helicity angle $\chi$
in $\tau\to\omega\pi\nu_\tau$.}
\begin{center}
\begin{tabular}{||c|c|c||}
\hline
$J^P$ & $L$ & $F(\cos\chi)$ \\  \hline\hline
$1^-$ & $1$ & $1-\cos^2\chi$ \\ \hline
$1^+$ & $0$ & $1$ 	      \\ \hline
$1^+ $ & $ 2$  &  $ 1+3\cos^2\chi$ \\ \hline
$0^-$  & $ 1$  &  $ \cos^2\chi$  \\ \hline
\end{tabular}
\end{center}
\end{table}

\begin{figure}[htb]
\centerline{\psfig{figure=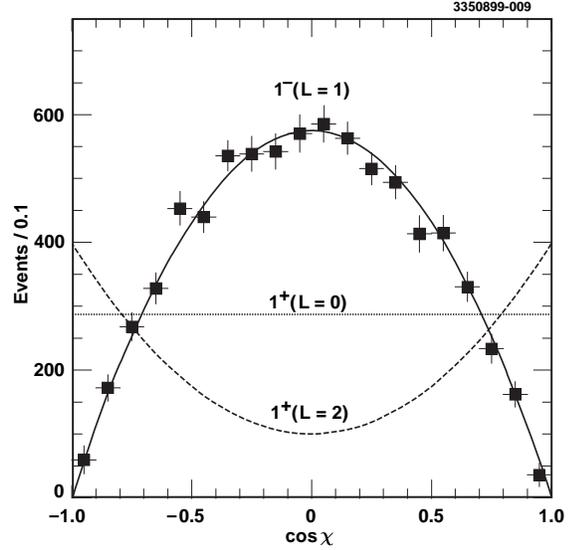,width=75mm}}
\caption[]{\label{fig:cleocoschi}
Fit to the $\cos\chi$ distribution
for the CLEO $\tau\to\omega\pi\nu_\tau$ data.
Black squares are the data, lines are the predictions for the different
$J^P$ contributions.
}
\end{figure}

The fit to the $\cos\chi$ distribution 
for the CLEO $\tau\to\omega\pi\nu_\tau$ data
(corrected for background and efficiency) is shown in
Fig.~\ref{fig:cleocoschi}.
There is no evidence for non-vector current contributions,
and CLEO sets the limit
\ben
N^{\omega\pi}\mbox{(non-vec)}
/ N^{\omega\pi}\mbox{(vector)} < 6.4\% 
\een
at 95\% CL,
to be compared with ALEPH's limit \cite{ref:alephrhopipi} of $<$ 8.6\%.

\ssection{\bmath $\tau\to 3\pi\pi^0\nu$ resonant structure}

We perform an unbinned maximum likelihood fit in the full
kinematical space of the $\tau\to 3\pi\pi^0\nu$ decay,
to extract a model-dependent description of its resonant structure.
We use the structure function approach to 
describe the production of the $4\pi$ system from tau decay,
averaging over the unseen neutrino:
\ben
|{\mathcal{M}}|^2 =\frac{G^2_F}{2}V^2_{ud}{ L^{\mu\nu}J_\mu J^*_\nu},
\een
\ben
f_S = \overline{L^{\mu\nu}J_\mu J^*_\nu}=2(M^2_{\tau}-q^2)\sum_{i=1}^{16}
\overline{L_i}W_i.
\een
The hadronic current $J^\mu$ is modeled in terms of
resonances:
\ben
J^{\mu} = \alpha_\omega f_\omega^\mu F_{\omega}(q) +   
         \sum_k \alpha_k f_k^\mu F_k(q),
\een
\begin{eqnarray}
F_k(q)&=&\beta_k^0 + \beta_k BW_{\rho}(q) + \nonumber \\
 &&\beta'_k BW_{\rho'}(q) + \beta''_k BW_{\rho''}(q),
\end{eqnarray}
where $k$ runs over substructure components of the model
(see below). The $\alpha$'s and $\beta$'s are fit parameters.
The background PDF is modeled using an empirical form derived from the 
CLEO data.

\begin{figure}[htb]
\centerline{\psfig{figure=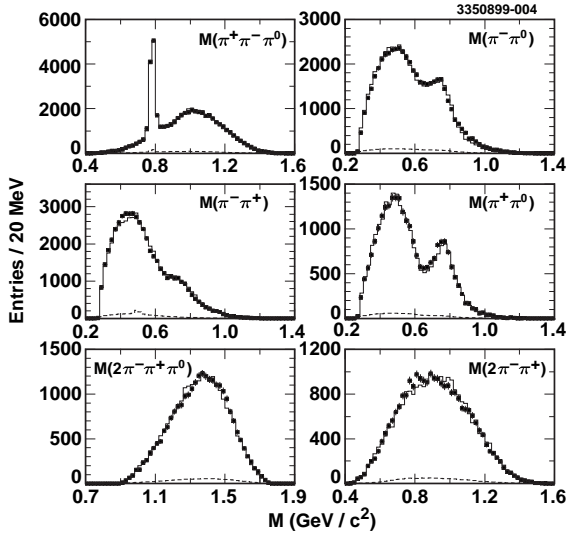,width=75mm}}
\caption[]{\label{fig:cleo4pifit}
Mass and sub-mass distributions in $\tau\to 3\pi\pi^0\nu$ decays
as measured by CLEO (data points).
The histograms are the results of a fit in the full
kinematical space of this decay, using Model 2 as described in the text.
}
\end{figure}

Four models were tried:
\begin{itemize}
 \item Model 1:  $\omega\pi$, $\rho\pi\pi$ and non-resonant $3\pi\pi^0$;
 \item Model 2:  $\omega\pi$ and $a_1\pi$;
 \item Model 3:  $\omega\pi$, $a_1\pi$, $\sigma\rho$ and $f_0(980)\rho$;
 \item Model 4:  $\omega\pi$, $a_1\pi$ and $\rho\pi\pi$.
\end{itemize}
The projections of the data, and the results of the fit to 
Model 2, are shown in Fig.~\ref{fig:cleo4pifit}.
There is clear evidence for contributions from
$\omega \pi^-$, $\rho^0 \pi^- \pi^0$,
$\rho^- \pi^+ \pi^-$, and $\rho^+ \pi^- \pi^-$.
The CLEO data prefer models containing at least
$\omega\pi$ and $a_1\pi$, in good agreement with results from 
the CMD-2 analysis \cite{ref:cmd24pi} of $e^+e^- \to 4\pi$.
The data do not rule out 
small contributions from modes like $\sigma \rho$,
$f_0 \rho$, or non-resonant$\rho\pi\pi$.

\ssection{Test of CVC in $4\pi$}

\begin{figure}[htb]
\centerline{\psfig{figure=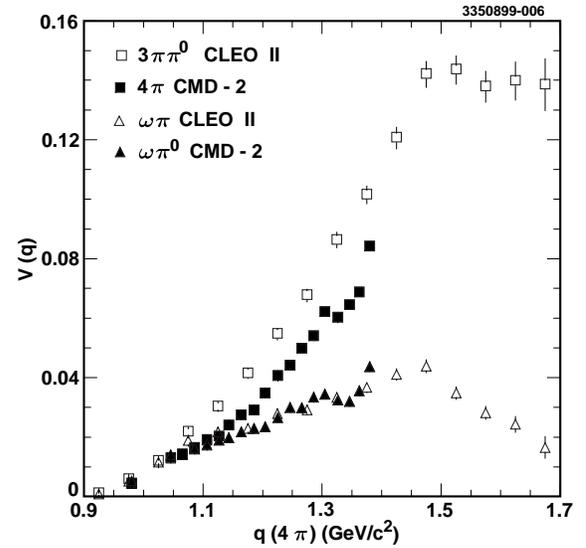,width=75mm}}
\caption[]{\label{fig:4picvc}
Comparison of the $4\pi$ and $\omega\pi$ spectral functions
measured (by CLEO) in $\tau\to 4\pi\nu_\tau$ decays with
the analogous ones measured (by CMD-2 \protect\cite{ref:cmd24pi}) 
in $e^+e^-$ annihilation.
}
\end{figure}

As a test of CVC, 
we can compare the $4\pi$ and $\omega\pi$ spectral functions
measured in $\tau\to 4\pi\nu_\tau$ charged current decays with 
the analogous ones measured (by CMD-2 \cite{ref:cmd24pi}) 
in $e^+e^-$ annihilation via the neutral EM current:
\begin{eqnarray}
V^{3\pi\pi^0}(q) & = & \frac{q^2}{4\pi^2\alpha^2}
\left[ \frac{1}{2} \sigma_{e^+e^- \to 2\pi^+2\pi^-}(q) \right. \nonumber \\
&& +\left. \sigma_{e^+e^- \to \pi^+\pi^-2\pi^0}(q) \right]\ ; \\
V^{\omega\pi}(q) & = & \frac{q^2}{4\pi^2\alpha^2} 
\sigma_{e^+e^- \to \omega\pi^0}(q)\ . 
\end{eqnarray}
This is done in Fig.~\ref{fig:4picvc}.

The CMD-2 data, which include a
15\% systematic error on the overall normalization,
show clear dominance of $a_1\pi$ and $\omega\pi$ in $e^+e^-\to 4\pi$.
We see that the shapes agree well between $\tau$ and $e^+e^-$ data.
However, the 
normalization of the $4\pi$ (non-$\omega\pi$) spectral functions
do not agree with one another. 
This is also seen in the total branching fraction
for $\tau\to 3\pi\pi^0\nu$ predicted \cite{ref:eidelman98}
from the (pre-CMD-2) $e^+e^-$ data using CVC:
\begin{eqnarray}
B(\tau\to 3\pi\pi^0\nu) =&& (4.22\pm 0.10)\% \\
CVC (excl.~CMD-2):      & &  (4.06\pm 0.25)\%.
\end{eqnarray}
This disagreement might be due to normalization errors,
other experimental errors, or a real violation of CVC.

\section{$5\pi$, $6\pi$, $7\pi$}

The decays of the $\tau$ to $5\pi\nu$, $6\pi\nu$, and $\ge 7\pi\nu$
all have small branching fractions:
\begin{eqnarray}
  \BR(2\pi^-\pi^+2\pi^0 \nu_\tau)  &= (5.3\pm 0.4)\times 10^{-3} \\
  \BR(3\pi^-2\pi^+      \nu_\tau)  &= (7.8\pm 0.6)\times 10^{-4} \\
  \BR(2\pi^-\pi^+3\pi^0 \nu_\tau)  &= (2.9\pm 0.8)\times 10^{-4} \\
  \BR(3\pi^-2\pi^+\pi^0 \nu_\tau)  &= (2.2\pm 0.5)\times 10^{-4} \\
  \BR(3\pi^-2\pi^+2\pi^0\nu_\tau)  &< 1.1\times 10^{-4} \\
  \BR(7\pi^\pm(\pi^0)   \nu_\tau)  &< 2.4\times 10^{-6}.
\end{eqnarray}

The limits on $\ge 7\pi\nu$ come from CLEO \cite{ref:cleo7pi}.
CLEO has no new results on $5\pi$ modes, but there are 
new results from CLEO \cite{ref:cleo6pi} on
$2\pi^-\pi^+3\pi^0$ and $3\pi^-2\pi^+2\pi^0$.

The decays $\tau^-\to (6\pi)^-\nu_\tau$ have three modes:
  \bi \ib $\tau^-\to 3\pi^- 2\pi^+ \pi^0\nu_\tau$
      \ib $\tau^-\to 2\pi^- \pi^+ 3\pi^0\nu_\tau$
      \ib $\tau^-\to  \pi^-       5\pi^0\nu_\tau$
  \ei
The mode $\tau^-\to  \pi^- 5\pi^0\nu_\tau$ has low efficiency
and large combinatoric background; CLEO has no signal or results
on this mode.

These decays all have very complex sub-structure.
We expect significant resonant substructure 
($\rho$'s, $\omega$'s, $\eta$'s).
We expect the $(6\pi)^-$ system to be dominantly 
produced via the vector current, although there can be
axial vector current contributions from $(3\pi)^-\eta$,
$\eta\to 3\pi$.

From data on the $I=1$ component of 
$e^+e^-\to 6\pi$, CVC predicts~\cite{ref:eidelman97}
\ben
v_1^{6\pi}(q^2) = \frac{q^2}{4\pi^2\alpha^2} 
            \sigma_{I=1}(e^+e^-\to 6\pi) ,
\een
\ben
\BR(\tau^-\to (6\pi)^-\nu_\tau)\ge (1.23\pm 0.19)\times 10^{-3}.
\een

\ssection{Isospin in $\tau^-\to (6\pi)^-\nu_\tau$}

As a first step in describing the resonant substructure in
$6\pi$ states, we can classify them by their
isospin content, with ``partitions'' (triplets of numbers $n_1,n_2,n_3$):
\bi
\item[\hbox to 9.5mm{$n_3$\hfill}] = number of $3\pi$ isoscalar systems ($\omega$) ,
\item[\hbox to 9.5mm{$n_2-n_3$\hfill}] = number of $2\pi$ isovector systems ($\rho$), 
\item[\hbox to 9.5mm{$n_1-n_2$\hfill}] = number of $\pi$ isovector systems.
\ei
For example, $(3,2,1) = (\pi\rho\omega)$.
The contributions of each partition to final states are given by:
{
\begin{eqnarray}
\Gamma(\pi^- 5\pi^0) &=& \frac{9}{35}\Gamma(4\pi\rho) \\
\Gamma(2\pi^- \pi^+ 3\pi^0) &=& \frac{2}{7}\Gamma(4\pi\rho)
   +\frac{1}{5}\Gamma(3\pi\omega) \nonumber \\
 &&+\frac{4}{5}\Gamma(3\rho)
   +\frac{1}{2}\Gamma(\pi\rho\omega) \\
\Gamma(3\pi^- 2\pi^+ \pi^0) &=& \frac{16}{35}\Gamma(4\pi\rho)
   +\frac{4}{5}\Gamma(3\pi\omega) \nonumber \\
 &&+\frac{1}{5}\Gamma(3\rho)
   +\frac{1}{2}\Gamma(\pi\rho\omega) .
\end{eqnarray}
}

\begin{figure}[htb]
\centerline{\psfig{figure=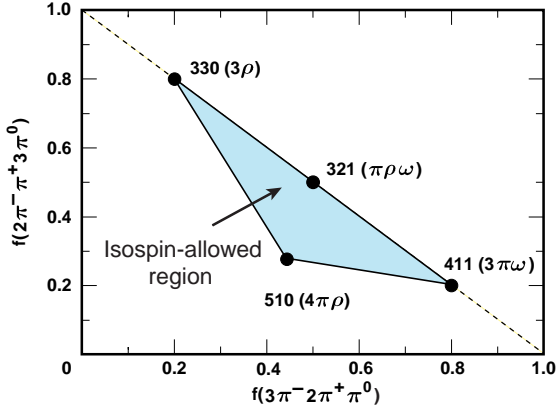,width=75mm}}
\caption[]{\label{fig:6piiso}
Isospin-allowed region in the space of $6\pi$ partial rate
fractions, as described in the text.
}
\end{figure}

Isospin conservation constrains the partial rates
{
    \begin{eqnarray}
     f(2\pi^- \pi^+ 3\pi^0) &=& \Gamma(2\pi^- \pi^+ 3\pi^0)/\Gamma(6\pi) \\
     f(3\pi^- 2\pi^+ \pi^0) &=& \Gamma(3\pi^- 2\pi^+ \pi^0)/\Gamma(6\pi) \\
     f(\pi^- 5\pi^0) &=& 1 - f(2\pi^- \pi^+ 3\pi^0) \nonumber \\
                     & & - f(3\pi^- 2\pi^+ \pi^0)
    \end{eqnarray}
}
to lie inside the space illustrated in Fig.~\ref{fig:6piiso}.
CLEO can measure only the ratio
$f(2\pi^- \pi^+ 3\pi^0) / f(3\pi^- 2\pi^+ \pi^0)$.

Axial vector currents can produce $6\pi$ final states,
due primarily (one assumes) to intermediate states $(3\pi)^-\eta$,
$\eta\to 3\pi$.
Two mechanisms have been proposed to predict the rate for these decays.
A G-parity violating term, suppressed by $(m_u^2-m_d^2)/(m_u^2+m_d^2)$
yields the prediction \cite{ref:pich87}
\ben
\BR(\pi^-2\pi^0\eta\nu_\tau) \simeq 
     \BR(2\pi^-\pi^+\eta\nu_\tau) \simeq 1.2\times 10^{-6}.
\een
An anomalous Wess-Zumino term without $(m_u^2-m_d^2)$ suppression
yields the predictions \cite{ref:binganli97}
\ben
\BR(f_1\pi^-\nu_\tau) = 2.9\times 10^{-4}
\een
\ben
\BR(\pi^-\rho^0\eta\nu_\tau \to 2\pi^-\pi^+\eta\nu_\tau)
         = 2.9\times 10^{-4}
\een

\ssection{CLEO results for $\tau^-\to 6\pi\nu_\tau$}

\begin{figure}[htb]
\centerline{\psfig{figure=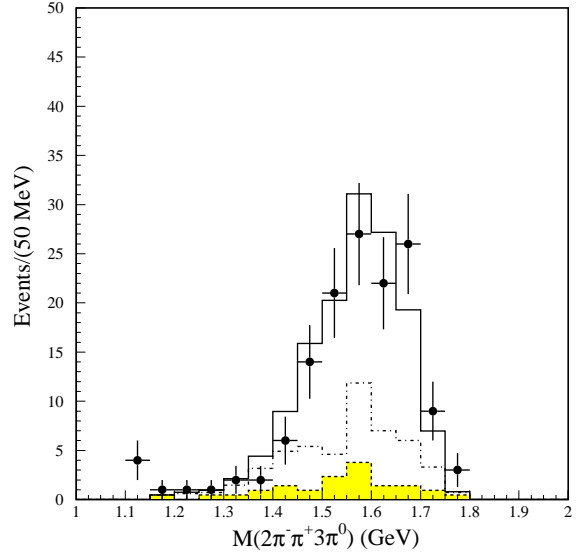,width=75mm}}
\caption[]{\label{fig:cleo3pi3pi0}
The $m(6\pi)$ spectrum from 
$\tau^-\to 2\pi^-\pi^+ 3\pi^0\nu_\tau$ as measured by CLEO.
The points are data. 
The hatched histogram is an estimate of the
hadronic background.
The dotted histogram is
background from tau decays,
and the solid histogram is the prediction for the signal plus backgrounds.
}
\end{figure}

\begin{figure}[htb]
\centerline{\psfig{figure=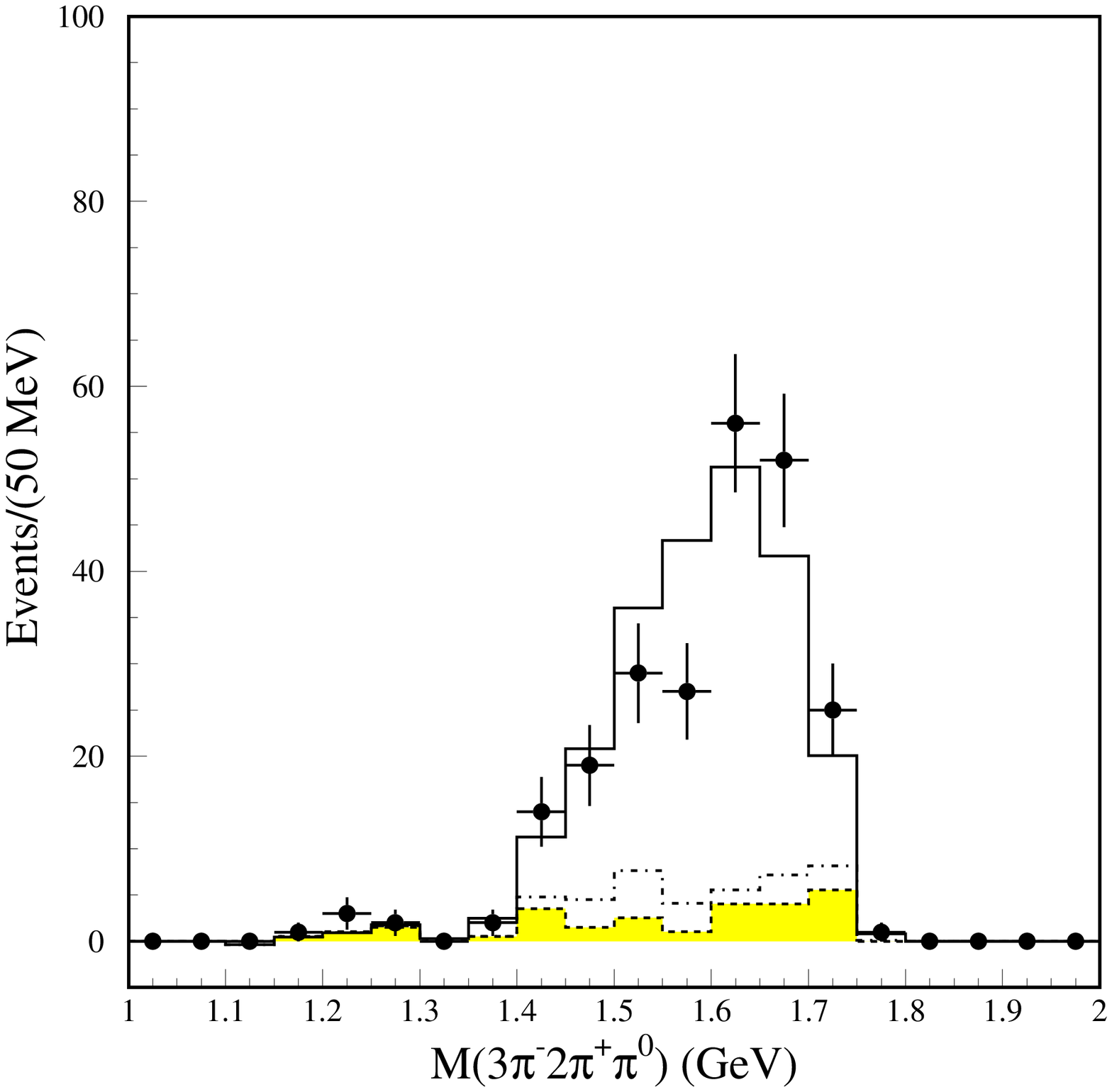,width=75mm}}
\caption[]{\label{fig:cleo5pipi0}
The $m(6\pi)$ spectrum from 
$\tau^-\to 3\pi^-2\pi^+ \pi^0\nu_\tau$ as measured by CLEO.
The points are data. 
The hatched histogram is an estimate of the
hadronic background.
The dotted histogram is
background from tau decays,
and the solid histogram is the prediction for the signal plus backgrounds.
}
\end{figure}

Using the full CLEO II and CLEO II.V data set,
corresponding to $12.3\times 10^6$ produced $\tau^+\tau^-$ pairs,
we reconstruct $(139\pm 12)$ events in the mode
$\tau^-\to 2\pi^-\pi^+ 3\pi^0\nu_\tau$,
with a background of 36\% from other tau decays and from
hadronic events.
The $6\pi$ mass spectrum is shown in Fig.~\ref{fig:cleo3pi3pi0}.
We measure \cite{ref:cleo6pi} a branching fraction
\ben
\BR(2\pi^-\pi^+ 3\pi^0\nu_\tau) = 
       (2.2\pm 0.3\pm 0.4)\times 10^{-4}.
\een
We reconstruct $(231\pm 19)$ events in the mode
$\tau^-\to 3\pi^-2\pi^+ \pi^0\nu_\tau$,
with a background of 20\% from other tau decays and from
hadronic events.
The $6\pi$ mass spectrum is shown in Fig.~\ref{fig:cleo5pipi0}.
We measure \cite{ref:cleo6pi} a branching fraction
\ben
\BR(3\pi^-2\pi^+ \pi^0\nu_\tau) =
       (1.7\pm 0.2\pm 0.2)\times 10^{-4}.
\een
These results compare well with previous results from
CLEO \cite{ref:cleo3pi3pi0,ref:cleo5pipi0},
ALEPH \cite{ref:aleph5pipi0},
and OPAL \cite{ref:opal5pipi0},
but with smaller errors.

\ssection{Resonant substructure in $\tau^-\to 6\pi\nu_\tau$}

\begin{figure}[htb]
\centerline{
\psfig{figure=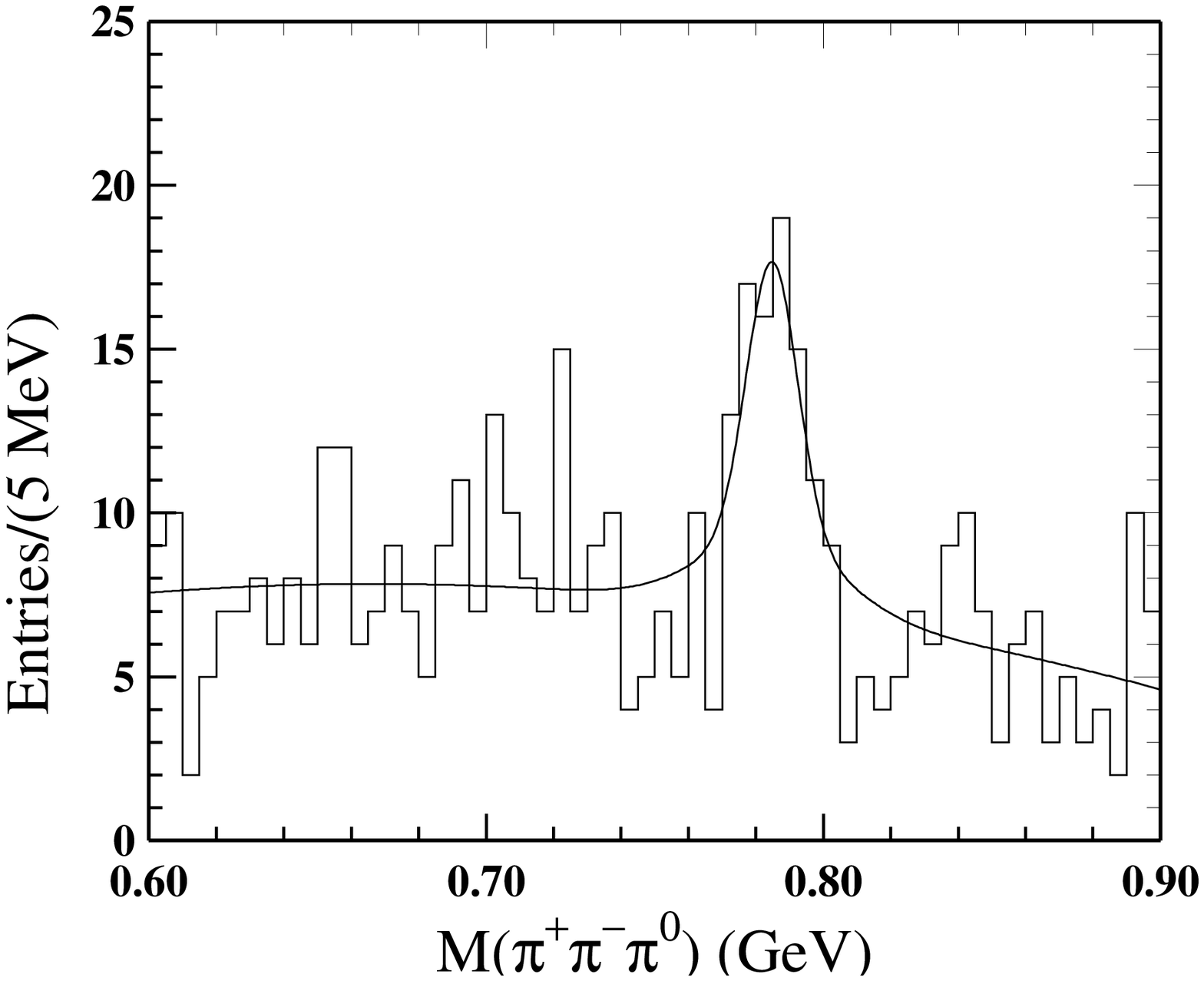,width=37mm}
\psfig{figure=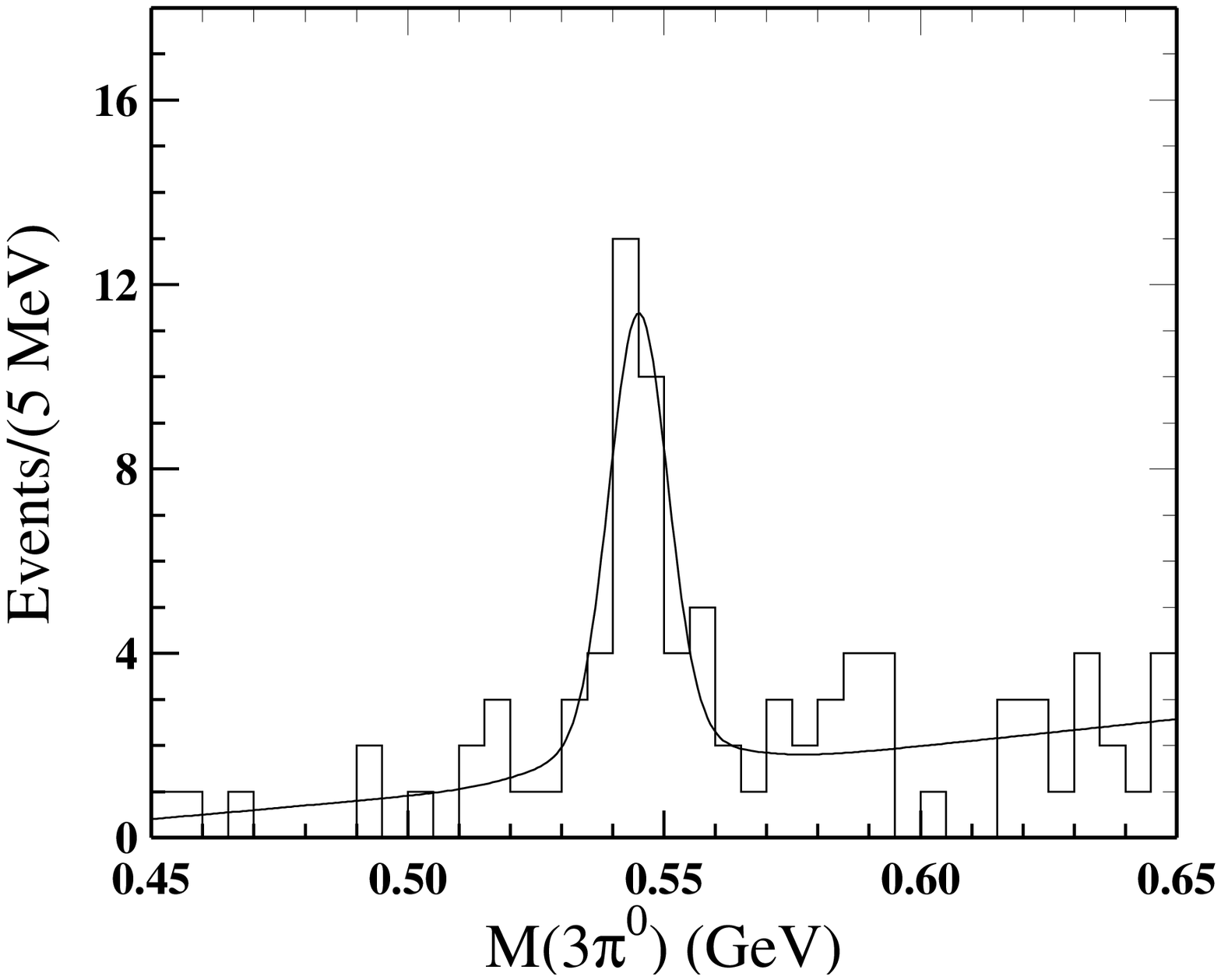,width=37mm}
}
\caption[]{\label{fig:cleo6pisub}
Some sub-mass distributions in $\tau\to 6\pi\nu_\tau$.
Left: $m(\pi^+\pi^-\pi^0)$ in $\pi^-2\pi^0\omega\nu_\tau$,  $\omega\to\pi^+\pi^-\pi^0$
(6 entries/event).
Right: $m(3\pi^0)$ in $2\pi^-\pi^+\eta\nu_\tau$,    $\eta\to 3\pi^0$
(1 entry/event).
}
\end{figure}

Clear signals are seen, and branching fractions measured,
for the following decay chains:
\bi
\ib $\tau^-\to \pi^-2\pi^0\omega\nu_\tau$,  $\omega\to\pi^+\pi^-\pi^0$ 
\ei
\ben
 \BR( \pi^-2\pi^0\omega \nu_\tau) = (1.5\pm 0.4\pm 0.3)\times 10^{-4}
\een
\bi
\ib $\tau^-\to 2\pi^-\pi^+\eta\nu_\tau$,    $\eta\to 3\pi^0$ 
\ei
\ben
 \BR( 2\pi^-\pi^+\eta   \nu_\tau) = (2.9\pm 0.7\pm 0.5)\times 10^{-4}
\een
\bi
\ib $\tau^-\to \pi^-2\pi^0\eta  \nu_\tau$,  $\eta  \to\pi^+\pi^-\pi^0$ 
\ei
\ben
 \BR( \pi^-2\pi^0\eta   \nu_\tau) = (1.5\pm 0.6\pm 0.3)\times 10^{-4}
\een
\bi
\ib $\tau^-\to 2\pi^- \pi^+\omega\nu_\tau$, $\omega\to\pi^+\pi^-\pi^0$
\ei
\ben
 \BR( 2\pi^-\pi^+\omega \nu_\tau) = (1.2\pm 0.2\pm 0.1)\times 10^{-4}
\een
\bi
\ib $\tau^-\to 2\pi^- \pi^+\eta  \nu_\tau$, $\eta  \to\pi^+\pi^-\pi^0$
\ei
\ben
 \BR( 2\pi^-\pi^+\eta   \nu_\tau) = (1.9\pm 0.4\pm 0.3)\times 10^{-4}.
\een
Some representative sub-mass distributions
are shown in Fig.~\ref{fig:cleo6pisub}.
This constitutes the first observation of $\tau^-\to 2\pi^- \pi^+\omega\nu_\tau$,
and the first observations of $\tau^-\to 3\pi\eta  \nu_\tau$ in 
the $\eta  \to 3\pi$ decay modes.

\ssection{Separating vector and axial-vector contributions}

We can combine these $(3\pi)^-\eta \to (6\pi)^-$ results 
with measurements using $\eta\to\gamma\gamma$~\cite{ref:cleo3pieta}, to get:
\ben
\BR_{av}( 2\pi^-\pi^+\eta \nu_\tau) = (2.4\pm 0.5)\times 10^{-4},
\een
\ben
\BR_{av}( \pi^-2\pi^0\eta \nu_\tau) = (1.5\pm 0.5)\times 10^{-4}.
\een
The $(3\pi)^-\eta$ system has a rich substructure,
only beginning to be explored; for example,
it can arise through the decay chain~\cite{ref:cleo3pieta}
$f_1\pi$, $f_1\to a_0\pi$, $a_0\to \eta\pi$.

We can then subtract these contributions from the total
$\tau\to 6\pi\nu_\tau$ decay rate;
what's left is presumed to be from the vector current only:
\ben
\BR_V( 2\pi^-  \pi^+ 3\pi^0\nu_\tau) = (1.1\pm 0.4)\times 10^{-4},
\label{eqn:BV3pi3pi0}
\een
\ben
\BR_V( 3\pi^- 2\pi^+  \pi^0\nu_\tau) = (1.1\pm 0.2)\times 10^{-4}.
\label{eqn:BV5pipi0}
\een

The vector to $6\pi$ decay rate $\BR_V$
is consistent with being saturated by contributions from $(3\pi)^-\omega$.

\begin{figure}[htb]
\centerline{\psfig{figure=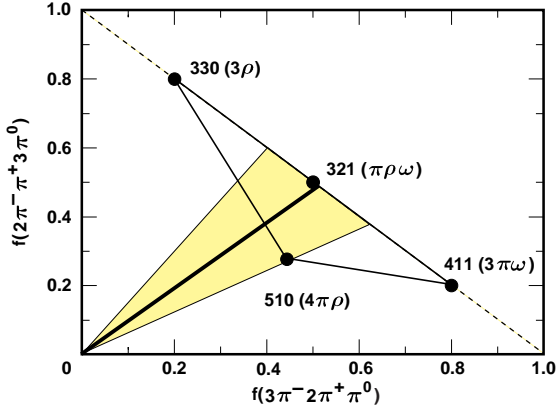,width=75mm}}
\caption[]{\label{fig:6piisocon}
Constraint from CLEO data on the allowed
region in the space of $6\pi$ partial rate
fractions; 
see Fig.~\protect\ref{fig:6piiso}.
}
\end{figure}

From the ratio of these branching fractions, we can constrain the region
of the isospin plane, as shown in Fig.~\ref{fig:6piisocon}.
Because the rate for $(6\pi)^-\to \pi^- 5\pi^0$ is unknown,
we cannot test whether the result lies within the 
isospin-allowed region.

\ssection{CVC predictions for $\tau\to (6\pi)^-\nu_\tau$}

The CLEO results on the vector part of the $\tau\to 6\pi\nu_\tau$
branching fractions, equations \ref{eqn:BV3pi3pi0} and \ref{eqn:BV5pipi0}, 
can be compared with the predictions 
from $e^+e^-\to 6\pi$ using CVC \cite{ref:eidelman97}:
\begin{eqnarray}
\BR_V( 2\pi^-  \pi^+ 3\pi^0\nu_\tau) &\ge& (2.5\pm 0.4)\times 10^{-4} \\
\BR_V( 3\pi^- 2\pi^+  \pi^0\nu_\tau) &\ge& (2.5\pm 0.4)\times 10^{-4} \\
\BR_V( (6\pi)^-            \nu_\tau) &\ge& (12.3\pm 1.9)\times 10^{-4} 
\end{eqnarray}

The CVC predictions are significantly higher than the CLEO results.
This may be due to an underestimate of the 
$I=0$ contributions to $\sigma(e^+e^-\to 6\pi)$.

\section{$\tau^-\to \nu_\tau\eta (n\pi)^-$}

The decay 
$\tau^-\to \eta\pi^-\nu_\tau $ is forbidden by G-parity;
the upper limit at 95\%\ CL
on this decay rate from CLEO \cite{ref:cleoetapi} is
\ben
\BR(\eta\pi^-\nu_\tau ) < 1.4\times 10^{-4}.
\een

The decay $\tau^-\to \eta\pi^-\pi^0\nu_\tau $ proceeds via the 
Wess-Zumino chiral anomaly; the branching fraction measured by CLEO \cite{ref:cleoetapi} 
\ben
\BR(\eta\pi^-\pi^0\nu_\tau ) = (1.7\pm 0.3)\times 10^{-3},
\een
is in good agreement with predictions.
However, the Wess-Zumino Lorentz structure has not been definitively 
established for this decay.

CLEO sees $\tau^-\to \eta(3\pi)^-\nu_\tau $ in two modes 
using $\eta\to\gamma\gamma$ \cite{ref:cleo3pieta},
as well as in the $6\pi$ modes reported above.
There is rich substructure in these modes, only beginning to be explored.
For example, there is evidence of $f_1$ production \cite{ref:cleo3pieta}:
$f_1\pi$, $f_1\to a_0\pi$, $a_0\to \eta\pi$.

The $SU(3)_f$-violating decay $\tau^-\to \eta K^-\nu_\tau$ has been 
seen by CLEO \cite{ref:cleoetak} at the rate
\ben
\BR(\eta K^-\nu_\tau ) = (2.6\pm 0.5)\times 10^{-4}.
\een

\ssection{$\tau^-\to K^{*-}\eta\nu_\tau$}

The decays $\tau^-\to K^{*-}\eta\nu_\tau$ were searched for in
the CLEO II sample of $\approx 4.3\times 10^6$ 
$\tau^+\tau^-$ pairs \cite{ref:cleoksteta}.
In the $\tau^-\to K_s \pi^-\eta\nu_\tau$ mode, 
13 events were observed, with an expected background of 1 event,
yielding a product branching fraction:
\begin{eqnarray}
\BR(\tau^-\to K^{*-}\eta\nu_\tau)\times
   \BR(K^{*-} \to K_s \pi^-) =&& \nonumber \\
 (1.18\pm 0.38\pm 0.12)\times 10^{-4}. &&
\end{eqnarray}
In the $\tau^-\to K^- \pi^0\eta\nu_\tau$ mode,
12 events were observed, with an expected background of 1 event,
yielding a product branching fraction:
\begin{eqnarray}
\BR(\tau^-\to K^{*-}\eta\nu_\tau)\times
   \BR(K^{*-} \to K^- \pi^0) = && \nonumber \\
 (0.69\pm 0.36\pm 0.28)\times 10^{-4}. &&
\end{eqnarray}

\begin{figure}[htb]
\centerline{
\epsfxsize=75mm\leavevmode\epsfbox[96 264 516 718]{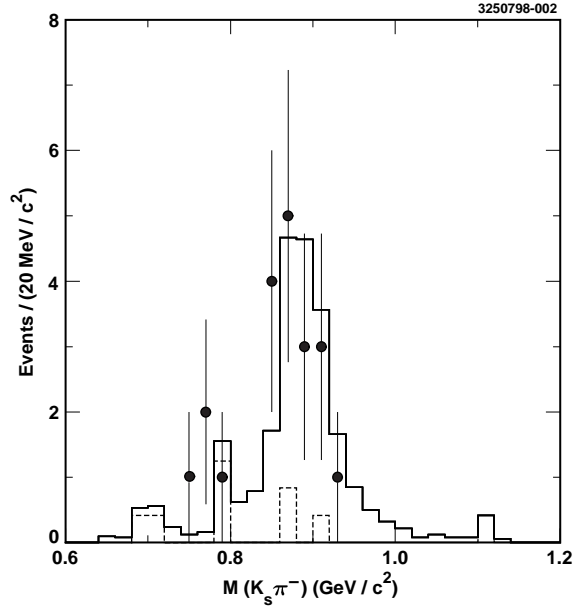}
}
\caption[]{\label{fig:kstareta}
The $M(K_S\pi^\mp)$ spectrum from 
$\tau^\mp\to K_S\pi^\mp\eta\nu_\tau$ events observed by CLEO,
showing evidence for $K^{*\mp}\to K_S\pi^\mp$.
The points are the data, dashed histogram is the expected background,
and solid histogram is the expected signal plus background.
}
\end{figure}

Combining the two modes, we obtain
\ben
\BR(K^{*-}\eta\nu_\tau) = (2.90\pm 0.80\pm 0.42)\times 10^{-4} ,
\een
and the mass spectrum shown in Fig.~\ref{fig:kstareta}.
This is the first observation of this decay mode,
and it is reasonable agreement with 
the prediction of $\sim 1\times 10^{-4} $ from Ref.~\cite{ref:binganli97}.

\section{Summary and conclusions}

We have presented recent results on the structure of the
hadronic systems in
$\tau\to 2\pi\nu_\tau$, $3\pi\nu_\tau$,  $4\pi\nu_\tau$, $6\pi\nu_\tau$,
and modes containing $\eta$ mesons.
There are also new results from CLEO on
$\tau\to K h\pi(\pi^0)\nu_\tau$ \cite{ref:cleokhh} and
$\tau\to K^- \pi^+\pi^-\nu_\tau$ \cite{ref:cleokpp},
which we have no room to report on here.

There are several apparent
`discrepancies'  between $\tau$  and $e^+e^-$ data.
This may be due to normalization problems,
other experimental errors, or a real violation
of CVC, which is expected at some level.
We need a better understanding of 
the applicability of CVC, to resolve these discrepancies.

The rich structure in multi-meson systems can certainly
be further elucidated.
It is clear that semi-hadronic tau lepton decay 
can be a powerful and unique probe of light hadronic systems.
The field is still very much driven by experiment.
It is hoped that the data will provide
stimulation for deeper theoretical work in this difficult field.

The data are also useful for studying other aspects of the
Standard Model, such as the 
tau neutrino helicity $h_{\nu_\tau}$,
the tau neutrino mass,
the running of the strong coupling constant $\alpha_s(m_\tau^2)$,
and contributions to vacuum polarization 
from low energy hadronic physics
relevant for predicting 
the muon anomalous magnetic moment $a_\mu^{had}$
and the running EM coupling $\alpha_{em}(M_Z)$.

We can expect more interesting results on low energy meson dynamics
in semi-hadronic tau decay, using high-statistics measurements
from B factories.

\section{Acknowledgements}
This submission represents the work of many CLEO collaborators,
supported by the US Department of Energy, US National Science Foundation,
and other sources. The author would like to thank the organizers
for a very stimulating and enjoyable conference in the beautiful city of Victoria.

\end{document}